\documentclass[twocolumn]{aastex631}
\usepackage{amsmath}
\usepackage{natbib}

\newcommand{\hi}{\mbox{H{\small I}}}

\shorttitle{VERTICO: Effects of Galaxy Environment in Virgo Galaxies}
\shortauthors{Villanueva et al.}
\graphicspath{{./}{figures/}}
\begin{document}

\title{VERTICO IV: Environmental Effects on the Gas Distribution and Star Formation Efficiency of Virgo Cluster Spirals}

\author[0000-0002-5877-379X]{Vicente Villanueva}
\affiliation{Department of Astronomy, University of Maryland, College Park, MD 20742, USA}

\author[0000-0002-5480-5686]{Alberto D. Bolatto} 
\affiliation{Department of Astronomy, University of Maryland, College Park, MD 20742, USA}

\author{Stuart Vogel}
\affiliation{Department of Astronomy, University of Maryland, College Park, MD 20742, USA}

\author[0000-0003-1845-0934]{Tobias Brown}
\affiliation{Herzberg Astronomy and Astrophysics Research Centre, National Research Council of Canada, 5071 West Saanich Rd, Victoria, BC, 8 V9E 2E7, Canada}

\author[0000-0001-5817-0991]{Christine D. Wilson}
\affiliation{Department of Physics \& Astronomy, McMaster University, 1280 Main Street W, Hamilton, ON, L8S 4M1, Canada}

\author{Nikki Zabel}
\affiliation{Department of Astronomy, University of Cape Town, Private Bag X3, Rondebosch 7701, South Africa}

\author[0000-0002-1768-1899]{Sara Ellison}
\affiliation{Department of Physics \& Astronomy, University of Victoria, PO Box 1700 STN CSC, Victoria, BC, V8W 2Y2, Canada}

\author[0000-0003-1908-2168]{Adam R. H. Stevens}
\affiliation{International Centre for Radio Astronomy Research, The University of Western Australia, 35 Stirling Highway, Crawley WA 6009, Australia}

\author[0000-0002-9165-8080]{Mar\'ia Jes\'us Jim\'enez Donaire}
\affiliation{Observatorio Astron\'omico Nacional (IGN), C/Alfonso XII, 3, E-28014 Madrid, Spain}
\affiliation{Centro de Desarrollos Tecnol\'ogicos, Observatorio de Yebes (IGN), E-19141 Yebes, Guadalajara, Spain}

\author[0000-0002-0956-7949]{Kristine Spekkens}
\affiliation{Royal Military College of Canada, P.O. Box 17000, Station Forces, Kingston, ON, K7K 7B4, Canada}

\author[0000-0003-0080-8547]{Mallory Thorp}
\affiliation{Department of Physics \& Astronomy, University of Victoria, PO Box 1700 STN CSC, Victoria, BC, V8W 2Y2, Canada}

\author[0000-0003-4932-9379]{Timothy A. Davis}
\affiliation{Cardiff Hub for Astrophysics Research \&\ Technology, School of Physics \&\ Astronomy, Cardiff University, Queens Buildings, Cardiff, CF24 3AA, UK}

\author[0000-0003-4722-5744]{Laura C. Parker}
\affiliation{Department of Physics \& Astronomy, McMaster University, 1280 Main Street W, Hamilton, ON, L8S 4M1, Canada}

\author[0000-0002-0692-0911]{Ian D. Roberts}
\affiliation{Leiden Observatory, Leiden University, P.O. Box 9513, 2300 RA Leiden, The Netherlands}

\author{Dhruv Bisaria}
\affiliation{Department of Physics, Engineering Physics and Astronomy, Queen's University, Kingston, ON K7L 3N6, Canada}

\author{Alessandro Boselli}
\affiliation{Aix Marseille Univ, CNRS, CNES, LAM, Marseille, F-13013 France}

\author[0000-0003-0080-8547]{Barbara Catinella}
\affiliation{International Centre for Radio Astronomy Research, The University of Western Australia, 35 Stirling Highway, Crawley WA 6009, Australia}
\affiliation{ARC Centre of Excellence for All Sky Astrophysics in 3 Dimensions (ASTRO 3D), Australia}

\author[0000-0003-1440-8552]{Aeree Chung}
\affiliation{Department of Astronomy, Yonsei University, 50 Yonsei-ro, Seodaemun-gu, Seoul, 03722, Korea}

\author[0000-0002-7422-9823]{Luca Cortese}
\affiliation{International Centre for Radio Astronomy Research, The University of Western Australia, 35 Stirling Highway, Crawley WA 6009, Australia}
\affiliation{ARC Centre of Excellence for All Sky Astrophysics in 3 Dimensions (ASTRO 3D), Australia}

\author[0000-0002-3810-1806]{Bumhyun Lee}
\affiliation{Korea Astronomy and Space Science Institute, 776 Daedeokdae-ro, Daejeon 34055, Republic of Korea}

\author[0000-0002-9405-0687]{Adam Watts}
\affiliation{International Centre for Radio Astronomy Research, The University of Western Australia, 35 Stirling Highway, Crawley WA 6009, Australia}

\correspondingauthor{Vicente Villanueva-Llanos}
\email{vvillanu@umd.edu}



\begin{abstract}

We measure the molecular-to-atomic gas ratio, $R_{\rm mol}$, and the star formation rate (SFR) per unit molecular gas mass, SFE$_{\rm mol}$, in 38 nearby galaxies selected from the Virgo Environment Traced in CO (VERTICO) survey. We stack ALMA $^{12}$CO(J=2-1) spectra coherently using \hi\ velocities from the VIVA survey to detect faint CO emission out to galactocentric radii $r_{\rm gal} \sim 1.2\,r_{25}$. We determine their scale-lengths for the molecular and stellar components and find a roughly 3:5 ratio between them compared to $\sim1$:1 in field galaxies, indicating that the CO emission is more centrally concentrated than the stars. We compute $R_{\rm mol}$ as a function of different physical quantities. While the spatially-resolved $R_{\rm mol}$ on average decreases with increasing radius, we find that the mean molecular-to-atomic gas ratio within the stellar effective radius $R_{\rm e}$, $R_{\rm mol}(r<R_{\rm e})$, shows a systematic increase with the level of \hi, truncation and/or asymmetry (\hi\ perturbation). Analysis of the molecular- and the atomic-to-stellar mass ratios within $R_{\rm e}$, $R^{\rm mol}_{\star}(r<R_{\rm e})$ and $R^{\rm atom}_{\star}(r<R_{\rm e})$, shows that VERTICO galaxies have increasingly lower $R^{\rm atom}_{\star}(r<R_{\rm e})$ for larger levels  of H$_{\rm I}$perturbation (compared to field galaxies matched in stellar mass), but no significant change in $R^{\rm mol}_{\star}(r<R_{\rm e})$. We also measure a clear systematic decrease of the SFE$_{\rm mol}$ within $R_{\rm e}$, SFE$_{\rm mol}(r<R_{\rm e})$, with increasingly perturbed \hi. Therefore, compared to galaxies from the field, VERTICO galaxies are more compact in CO emission in relation to their stellar distribution, but increasingly perturbed atomic gas increases their $R_{\rm mol}$ and decreases the efficiency with which their molecular gas forms stars.
\end{abstract}

\date{October 2022. Accepted for publication in ApJ}


\section{Introduction} 
\label{intro}

A major aim in modern astrophysics is to understand how the local physical conditions of the interstellar medium (ISM) that lead to the production of stars respond to environmental effects. That is, whether there are differences between galaxies residing in a low-density environment and those immersed in galaxy clusters. 

Galaxies are known to not be evenly distributed in the Universe. For low/intermediate stellar masses, gas-rich galaxies are mainly found in low-density environments, suggesting a strong interplay between environment, gas cycle, and star-formation activity \citep[e.g.,][]{Koribalski2004,Meyer2004}. High-density environments also tend to preferentially host red galaxies \citep[e.g.,][]{Oemler1974,Goto2003,Thomas2010}. Numerical simulations and observational evidence have shown that galaxies lose the ability to accrete gas from the cosmic web when they fall into a more massive halo \citep[e.g.,][]{Dressler1980,Dressler1997,Dekel&Birnboim2006,Cappellari2011,Behroozi2019,Wright2022}, resulting in quenching of the star-formation activity once their original gas is depleted.

Galaxy clusters are the largest bound structures in the Universe, containing large number of galaxies tied by the cluster dark matter halos. Seminal studies have proposed several environmental mechanisms that may contribute to the quenching of star-formation. They range from strangulation/starvation \citep[i.e., galaxies cease to accrete gas cosmologically, and they continue forming stars until their remaining disk gas is consumed;][]{Larson1980,Balogh&Morris2000}, ram pressure stripping \citep[hereafter RPS, i.e. the removal of gas by `winds' due to the hot intra cluster medium, ICM;][]{Gunn&Gott1972}, galaxy interactions \citep[e.g., galaxy harassment;][]{Moore1996,Boselli&Gavazzi2006,Smith2010}, to the increase of gas stability through morphological quenching (hereafter MQ; \citealt{Martig2009}). However, the interplay between these different mechanisms and their relative contribution to changes in the gas content are still not \emph{precisely} understood \citep[e.g.,][]{Cortese2021}.

Star-formation activity takes place in giant molecular clouds, GMCs \citep[e.g.,][]{Sanders1985,WongBlitz2002,Kennicutt2007,Bigiel2008,Bigiel2011,Leroy2013}. However, H$_2$ reservoirs depend on the extended \hi\ component \citep[e.g.,][]{Verheijen2001}, and the presence of atomic gas is thus important for sustaining the production of new stars over a long time scale. Observations reveal that the atomic gas can be strongly affected by high-density environments. Cluster galaxies typically contain less atomic gas than their field counterparts, and commonly show signs of truncation and perturbed \hi\ morphologies \citep[e.g.,][]{Haynes&Giovanelli1984,Chung2009,Yoon2017,Brown2017,Stevens&Brown2017,Watts2020a,Watts2020b,Molnar2022}. Although the molecular gas is closer to galaxy centers and more tightly bound than the \hi\ \citep[][]{Davis2013}, several studies show that the H$_{2}$ is also susceptible to significant variations due to environmental effects (e.g., \citealt{Fumagalli2009,Boselli2014a,Lee2017,Lee2018,Zabel2019,Lizee2021,Stevens2021,Brown2021,Zabel2022}). 

How the molecular gas is affected by environment, and how this impacts the star-formation activity in galaxy clusters, is a topic of current research. Several scenarios have been proposed; they span from molecular and atomic gas being disturbed and removed simultaneously, atomic gas being removed before the molecular gas, or even an enhancement in the efficiency of the atomic-to-molecular gas transition due to the compression of \hi\ by ram pressure \citep[e.g.,][]{Chung&Kim2014}. Although using a small sample of galaxies with an heterogeneous set of molecular gas data,  \cite{Boselli2014a} find a mild statistically significant correlation between the H$_2$ and \hi\ deficiencies in Virgo galaxies from the Herschel Reference Survey \citep[HRS;][]{Boselli2010}, supporting the hypothesis that they are simultaneously affected. \cite{Zabel2022} note that VERTICO galaxies with larger \hi\ deficiencies (an indicator of how poor in \hi\ individual galaxies are when compared to field galaxies of the same size and morphological type, \citealt{Haynes&Giovanelli1984,Warmels1986,Cayatte1990}; see also \citealt{Cortese2021} for a detailed description) also have more compact and steeper H$_2$ radial profiles. Although with significantly different physical properties than the Virgo cluster, \cite{Loni2021} find however a significant scatter in the global molecular-to-atomic ratios, $R_{\rm mol}$, in galaxies selected from the Fornax cluster, which suggests that the effects of environmental mechanisms on the atomic-to-molecular gas transition may not be straightforward \citep[e.g., ][]{Stevens2021}. 

In recent decades, a broad variety of studies on the physical conditions within local field galaxies have shown that, for the H$_2$-dominated regions of galaxy disks, the star formation rate (SFR) per unit molecular gas mass, the star formation efficiency of the molecular gas, ${\rm SFE}_{\rm mol} = \Sigma_{\rm SFR} / \Sigma_{\rm mol}$, does not vary strongly with radius \citep[except for the galaxy centers; e.g.,][]{Schruba2010,Leroy2013,Villanueva2021}. However, \cite{Koopmann2004} found an anticorrelation between the H$\alpha$ central concentration parameter and the normalized massive star-formation rate (${\rm NMSFR}=F_{\rm H\alpha}/F_{R}$, where $F_{\rm H\alpha}$ and $F_{R}$ are the H$\alpha$ and the $R$-band fluxes, respectively) in Virgo galaxies. Moreover, detailed studies of Virgo's NGC 4654 galaxy by \cite{Chung&Kim2014} show that, even though $R_{\rm mol}$ values in the north-west appear to be lower when compared to other regions with similar total gas surface density, both the atomic gas surface density, $\Sigma_{\rm atom}$, and the SFE$_{\rm mol}$ seem to be higher. They associate this effect with atomic gas being compressed, which consequently increases the molecular gas surface density, $\Sigma_{\rm mol}$. Similar results are also found by \cite{Zabel2020}, who note an enhancement of the efficiencies in low mass galaxies on their first infall into the Fornax cluster. They suggest that this is likely due gas compression by environmental effects (e.g., by RPS and tidal interactions). The relations between scale lengths might be therefore significantly different in cluster galaxies due to environmental effects. \cite{Chung2009} observed that in galaxies closer to the center of the Virgo cluster (and also in some galaxies at the outskirts), \hi\ disks are much smaller than optical disks. Similar results are also found by \cite{Reynolds2022}, who find \hi\, disks smaller than optical disks in Hydra I cluster galaxies selected from the Widefield ASKAP L-band Legacy All-sky Blind Survey \citep[WALLABY; ][]{Koribalski2020}. However, it is less known how much the cluster environment might change the {\it molecular} gas distribution \citep[e.g.,][]{Mok2017}. Molecular gas is distributed closer to galaxy centers than \hi\ and it is thus more tightly bound and more difficult to affect, and some removal mechanisms such as RPS may be considerably less effective on the denser molecular medium. 

According to \cite{Krumholz2009}, the \hi-to-H$_2$ transition takes place through gas condensation of the atomic gas when it reaches a critical surface density of $\Sigma_{\rm crit}\approx 10$ M$_\odot$ pc$^{-2}$. Therefore, a plausible scenario is that RPS effects may be helping the \hi\ to reach the critical column density. As result, the enhancement of the H$_2$ production in Virgo galaxies may correspond to a more efficient \hi-to-H$_2$ transition. Several studies have reported observational evidence consistent with these ideas \citep[e.g., ][]{Roediger2014,Cramer2020,Lizee2021}. For instance, \cite{Lizee2021} note that NGC 4654 shows some strongly compressed atomic gas that exceeds $\Sigma_{\rm crit}$. They also found a CO-to-H$_2$ conversion ratio a factor $\sim2$ higher than the Galactic value and SFE$_{\rm mol}$ values around $\sim1.5-2$ higher than the rest of the disk. 

Galaxy surveys can be very helpful to understand the ensemble tendencies of the star-formation activity and how these depend on the physical conditions of the molecular gas (e.g., the HERA CO Line Extragalactic Survey, HERACLES, \citealt{Leroy2008,Leroy2013}; the Herschel Reference Survey, HRS, \citealt{Boselli2010}; the James Clerk Maxwell Telescope Nearby Galaxies Legacy Survey, NGLS, \citealt{Wilson2012}; the CO Legacy Database for GALEX Arecibo SDSS Survey, COLD GASS, and the extended COLD GASS, xCOLD GASS, \citealt{Saintonge2011b,Saintonge2017}). In this work, we determine radial length scales, mass ratios, and star formation efficiencies of molecular gas in the Virgo Environment Traced in CO survey \citep[VERTICO; ][]{Brown2021}, using ALMA Compact Array observations and ancillary data available for 38 galaxies with low inclinations. VERTICO is an ALMA Large Program designed to investigate the effect of the cluster environment on galaxies by mapping the star-forming molecular gas in 51 galaxies selected from the Virgo cluster. Since galaxy clusters are natural laboratories to test star-formation quenching processes due to environmental mechanisms, VERTICO gives us a unique opportunity to study not only their impact on molecular gas disturbances at sub-kiloparsec scales, but also to analyze how these processes affect the efficiency of the atomic-to-molecular gas transition and the star formation activity. 

This paper is organized as follows: Section \ref{S2_Observations} presents the main features of the VERTICO survey and the sample selection. In Section \ref{S4_Methods} we explain the methods applied to analyze the data and the equations used to derive the physical quantities. Finally, in Section \ref{S5_Results} we present our results and discussion, and Section \ref{S6_Conclusion} we summarize the main conclusions of this work.

\section{DATA PRODUCTS}
\label{S2_Observations}

One of the main advantages of carrying out a systematic analysis on the VERTICO survey is the vast ancillary data gathered by studies of Virgo cluster galaxies. The VERTICO sample selection and data reduction are described in detail in \cite{Brown2021}; here we summarize the main features.

\subsection{The VERTICO Survey}
\label{The_sample}

We use molecular gas data from VERTICO\footnote{https://www.verticosurvey.com}, which maps the CO(2-1) emission in 51 late-type spirals galaxies selected from the VIVA \hi\ survey \citep[][]{Chung2009}. The galaxies were observed with the ALMA Morita Array, including total power observations. Out of its 51 sources, 15 galaxies are taken from the ALMA archive \citep{Cramer2020,Leroy2021a}. VERTICO contains a broad diversity of galaxies experiencing different environmental effects, with stellar masses in the range $10^{8.3} \lesssim M_{\star}/M_{\odot} \lesssim 10^{11}$, and specific star formation rates, sSFR $=$ SFR$/M_\star$, of $10^{-11.5} \lesssim$ sSFR/yr$^{-1} \lesssim 10^{-9.5}$. VERTICO cover a variety of star formation properties in Virgo cluster galaxies, including normal (SFRs similar to the global median), enhanced (galaxies above 3 times the global median), anemic (galaxies with significantly low SFRs), and truncated galaxies (sharp cutoff in the star-forming disk), based on the spatial distribution of H$\alpha$ and $R$-band emission (see \S\, 2 in \citealt{Koopmann2004} for a detailed description of these categories).

VERTICO encompasses spectroscopic observations of the $J=2-1$ transition of $^{12}$CO and its isotopologues (i.e., $^{13}$CO(2-1) and $^{18}$CO(2-1)), as well as the 1-mm continuum. The galaxies were mapped using Nyquist-sampled mosaicking; while total power (TP) plus 7-m arrays were required for 25 galaxies, the rest of the observations were performed with 7-m array only. The archival data and raw visibilities were processed using the Common Astronomy Applications Software package v. 5.6 \citep[{\tt{CASA}};][]{McMullin2007}. The Compact Array data and the TP observations were combined using feathering and imaged with the PHANGS-ALMA Imaging Pipeline \citep[v. 1.0; ][]{LeroyPipe2021}. 

For the analysis performed in this work, we use the CO(2--1) 9\arcsec\, ($\sim750$ pc at Virgo cluster distance of 16.5 Mpc; \citealt{Mei2007}) datacubes with 10 km s$^{-1}$ channel width. Since NGC 4321 has a native angular resolution poorer than 9\arcsec\, ($\sim$10\arcsec), we used CO(2-1) datacubes at 15\arcsec\, ($\sim1$ kpc). When we compute the resolved molecular-to-atomic ratios (see \S\,\ref{Basic_equations}), we use the 15\arcsec\, CO(2--1) datacubes, which have been matched to VIVA's \hi \, angular resolution. The CO datacubes have a characteristic rms noise of $\sim$15 mJy beam$^{-1}$ at 5 km s$^{-1}$ (see \citealt{Brown2021} for more details).

\subsection{Ancillary Data and Data Selection}
\label{Database}

To complement the 9\arcsec \, CO(2--1) datacubes from VERTICO, we use the SFR surface density, $\Sigma_{\rm SFR}$, and stellar mass surface density, $\Sigma_{\star}$, derived from a combination of near UV and near/mid-infrared photometry. Specifically, we use the resolved star formation rate maps from GALEX and WISE photometry, which were derived by following the procedure laid out in \cite{Leroy2019}. All images are convolved from their native resolution to a 9\arcsec \, Gaussian beam with the \cite{Aniano2011} convolution kernels.  All Gaia DR2 stars within the image area are masked. Image backgrounds are estimated and subtracted with the {\tt{Astropy}} Background 2D function. SFR maps are constructed from a combination of GALEX NUV and WISE3 photometry as our obscured tracer in order to match the 9\arcsec \, resolution which is not possible with WISE4 (see Appendix of \citealt{Leroy2019} for more details). We apply a local WISE3-to-WISE4 colour correction to the WISE3 images by fitting a linear relationship between W3/W4 colour and galactocentric radius and then modify the WISE3 image on a pixel-by-pixel basis according to the galactocentric radius of each pixel and the expected W3/W4 ratio from the linear fit. For a more detailed description, see Jim\'enez-Donaire et al. (submitted). With GALEX NUV and colour-corrected WISE3 images for each galaxy, we then apply the NUV+WISE4 SFR calibration from \cite{Leroy2019} to derive spatially resolved SFR maps in units of M$_\odot$ kpc$^{-2}$ yr$^{-1}$. All the pixels in the maps where the S/N of the NUV or the WISE3 imaging is below 3 are masked. Stellar mass maps are derived from WISE1 photometry. For each pixel, the procedure determines the local mass-to-light ratio (at 3.4$\mu$m) using the WISE3 to WISE1 colour as an `sSFR-like' proxy and following the calibrations given in the Appendix of \cite{Leroy2019}. The WISE1 images are then combined with the derived mass-to-light ratios to produce resolved stellar mass maps in units of M$_\odot$ pc$^{-2}$. Both SFR and stellar maps at 9\arcsec \, are derived by assuming a Kroupa initial mass function (IMF; \citealt{Kroupa2001}). It would be interesting to verify that similar results are obtained using SFR estimators that respond on shorter timescales, such as H$\alpha$.

Optical inclination and position angles are obtained from fits to the Sloan Digital Sky Survey \citep[SDSS;][]{York2000,Alam2015} using $r$-band photometry. In order to measure the atomic gas content and velocities, we use 21-cm  moment-0 and moment-1 maps from the VLA Imaging survey of Virgo galaxies in Atomic gas \citep[VIVA; ][]{Chung2009}, which were re-imaged to $15\arcsec$ resolution to  match the resolution of VERTICO data \citep[][]{Brown2021}. Finally, the isophotal radius, $r_{25}$, is derived from the optical size of the major axis measured at 25 mag arcsec$^{-2}$ in the $B$-band, $r_{25}$ from the {\it The Third Reference Catalogue of Bright Galaxies} \citep[RC3; ][]{deVaucouleurs1991}.

To reduce the beam-smearing effects when deriving the molecular gas profiles, we select galaxies with inclinations $i\leq75^{\circ}$. Out of the 51 VERTICO galaxies, and rejecting the two nondetections of the survey (IC 3418 and VCC 1581; see Table 2 in \citealt{Brown2021}), we obtain a final subsample of 38 galaxies that fulfill the selection criteria.

\section{Methods}
\label{S4_Methods}

\subsection{Stacking of the CO Spectra}
\label{Stacking_CO}

\begin{figure}
\hspace{-0.175cm}
\includegraphics[width=8.75cm]{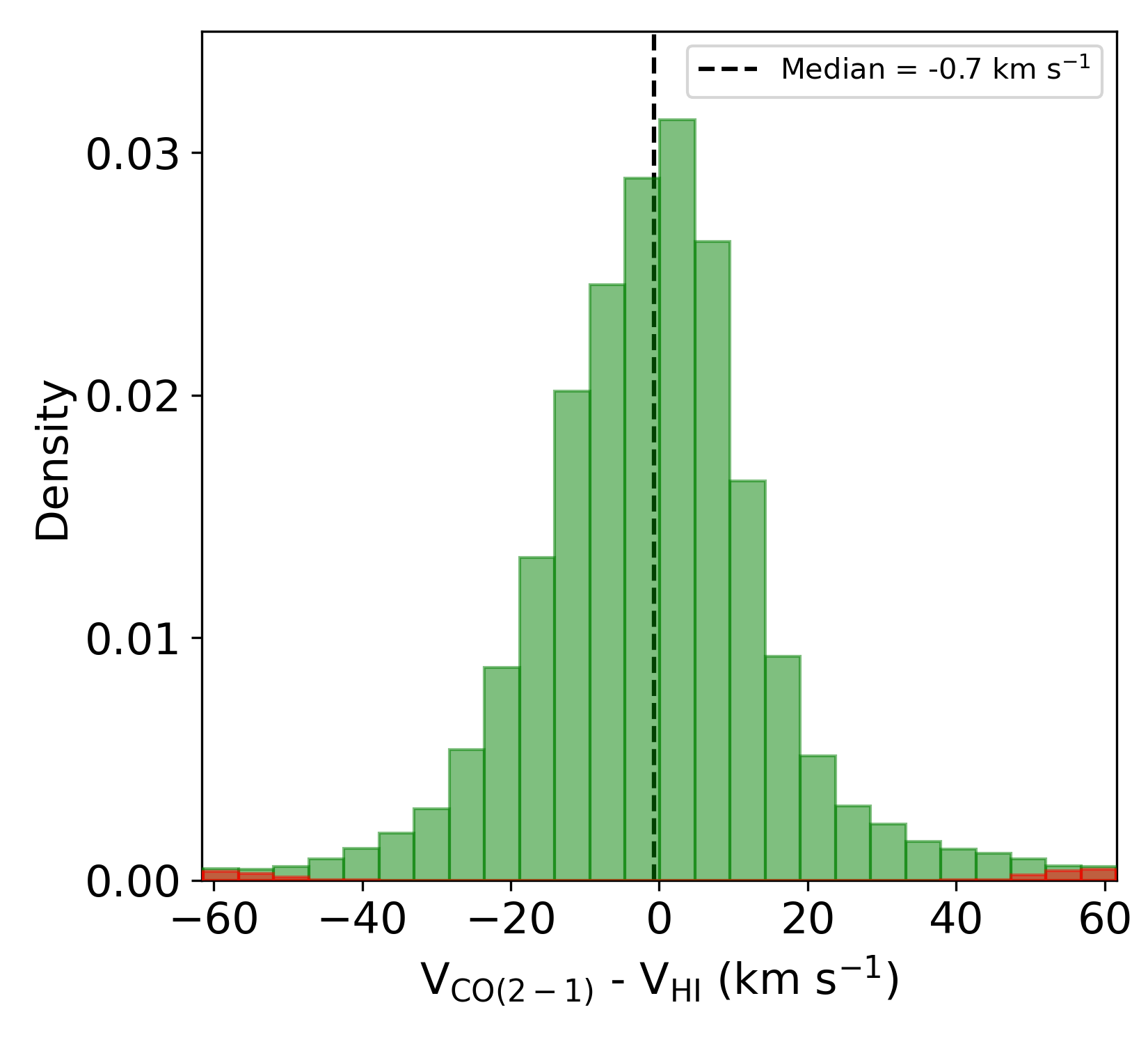}
\caption{ Distribution of offsets between the CO(2--1) and \hi\, velocities, $\Delta V= V_{\rm CO(2-1)} - V_{\rm HI}$, in spaxels within the 38 VERTICO galaxies analyzed here. The red bars correspond to spaxels with $\Delta V$ offsets that place CO outside the integration window (see text) for stacking CO. The vertical black-dashed line is the median value of $\Delta V= -0.7$ km s$^{-1}$. The figure shows that, on average, the differences between the CO(2--1) and \hi\, velocities are smaller than the integration window in most cases ($\sim98$\%).} 
\label{Vco_Vhi}
\end{figure}

\begin{figure*}
  \hspace{1cm}\includegraphics[width=15.cm]{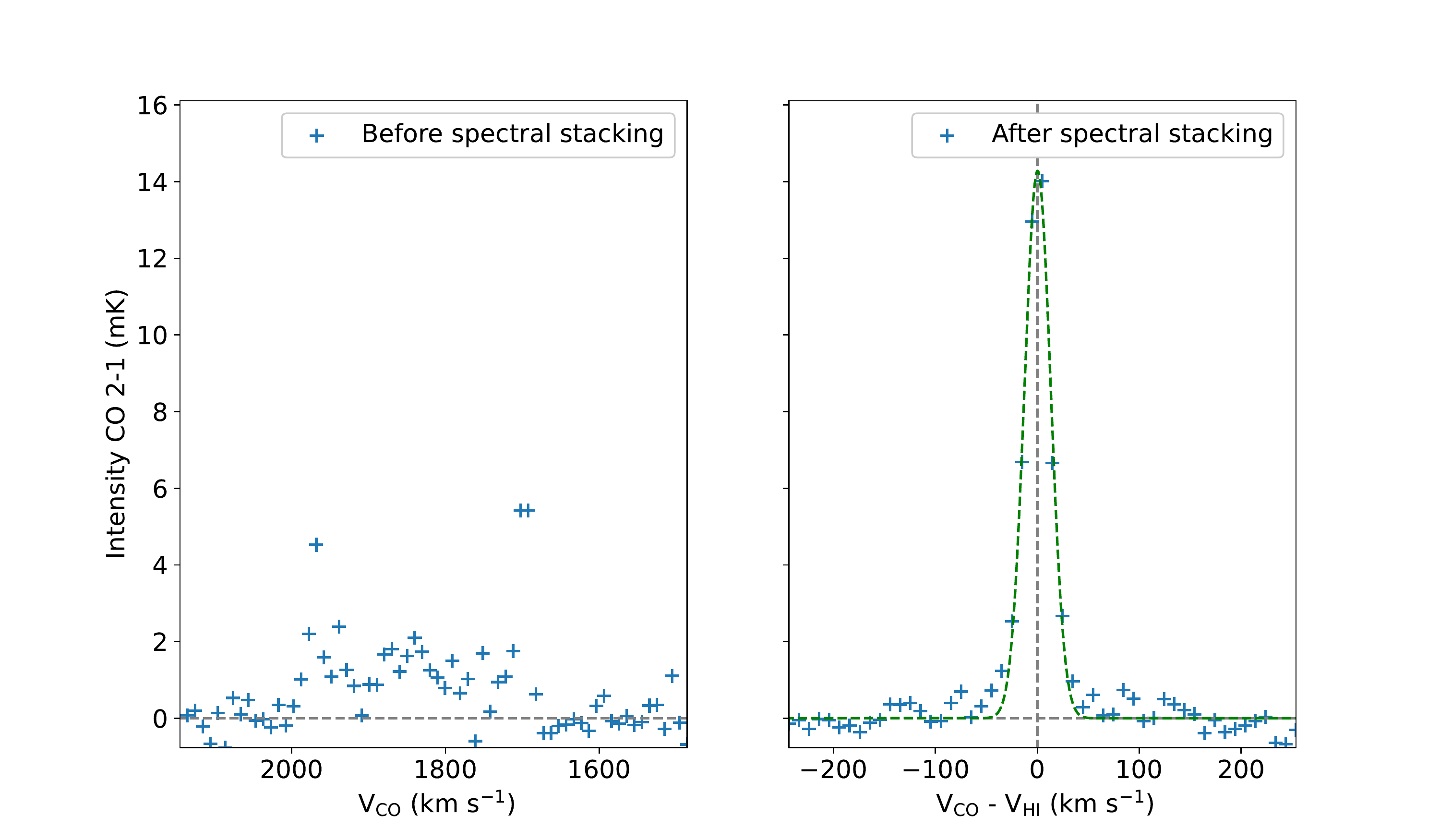}\llap{\shortstack{%
        \includegraphics[scale=.12]{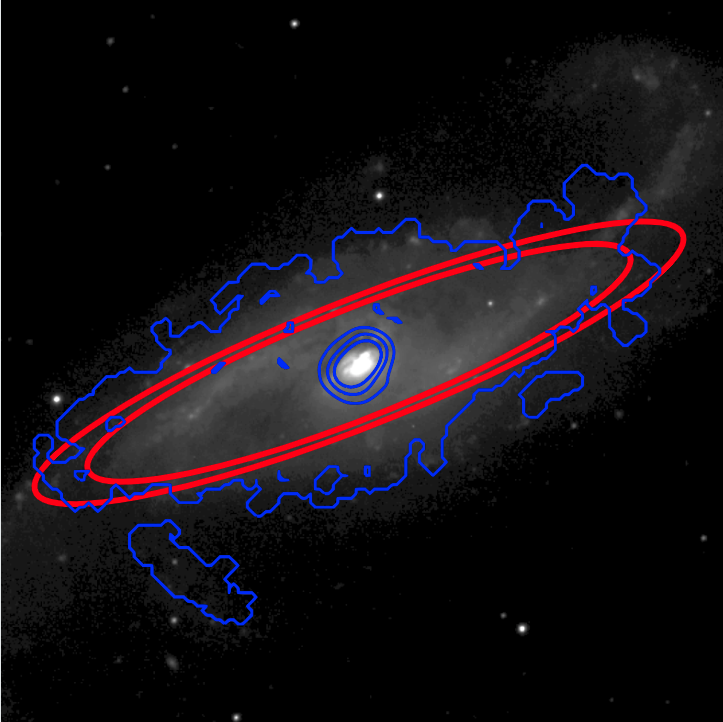}\\
        \rule{0ex}{1.5in}%
      }
  \rule{3.85in}{0ex}}
  \caption{Spectral stacking example. The average CO(2--1) spectrum within an annulus that spans from $0.6$ to $0.7$ $r_{25}$ in NGC $4536$ is shown. The left panel shows the average of all spectra in the annulus in the observed velocity frame. The inset panel includes the SDSS $r$-band image (background), CO(2--1) data (blue contours), and the annulus that spans from $0.6$ to $0.7$ $r_{25}$ (red ellipses). The right panel shows the average in the velocity frame relative to \hi\, along with the best Gaussian fit profile (green dashed line).}
 \label{Stacked_Example}
\end{figure*}

\begin{figure}
\includegraphics[width=8.cm]{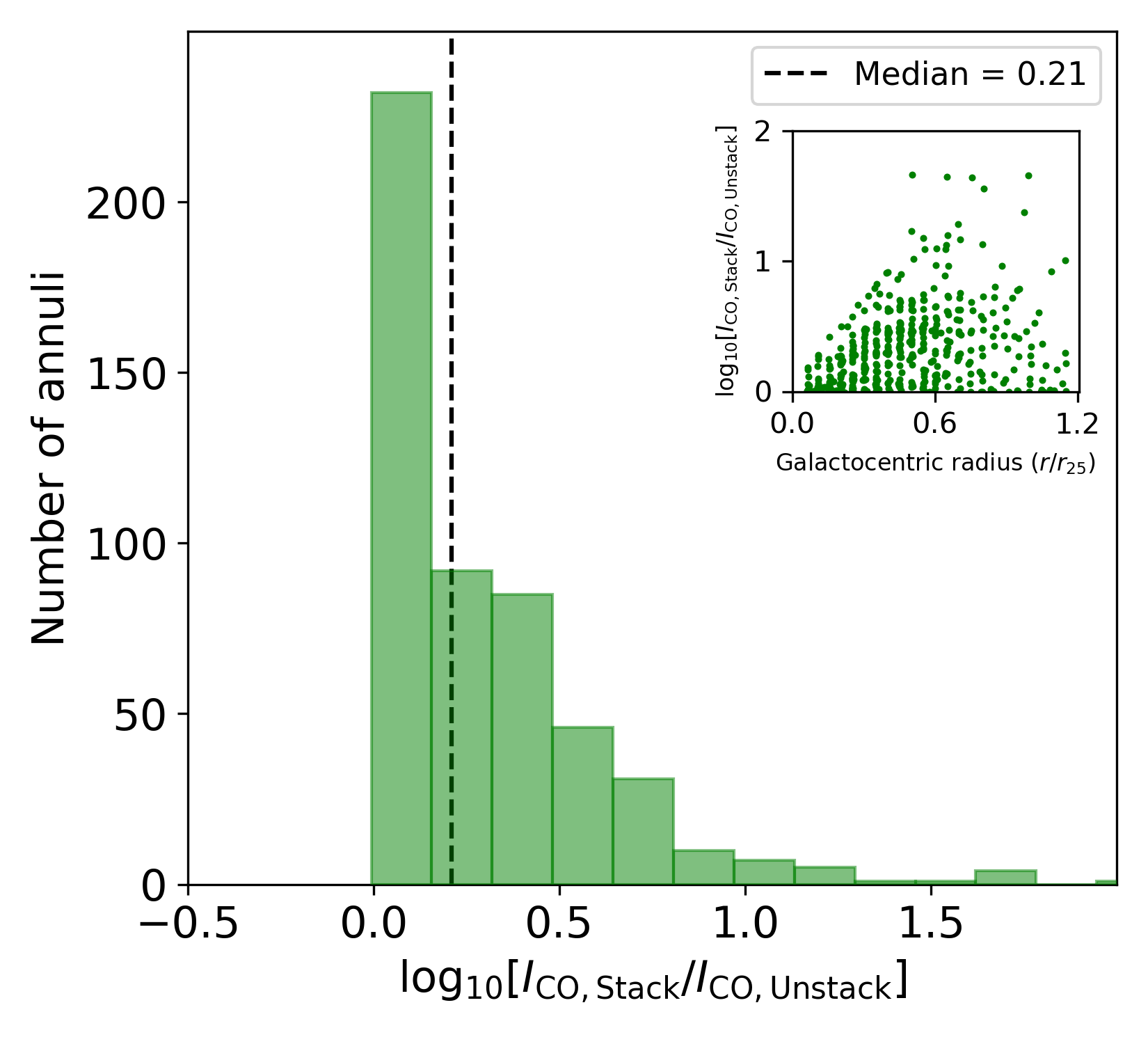}
\vspace{-0.2cm}
\caption{Ratio of the integrated CO(2--1) line intensity in an annulus after stacking to that before stacking. The vertical black-dashed line is the median value of log$_{10}[I_{\rm CO,Stack}/I_{\rm CO,Unstack}$] = $0.21$; this shows that, on average, stacking recovers $\sim60$\% more emission. On average, we are reaching a characteristic rms noise of 0.1 mK at 10 km s$^{-1}$, which corresponds to a sensitivity of $\sim0.1$ M$_{\odot}$ pc$^{-2}$. The inset panel compares  log$_{10}[I_{\rm CO,Stack}/I_{\rm CO,Unstack}$] vs  galactocentric radius, and shows that annuli with the most CO flux enhancement are at $r\gtrsim 0.5 r_{25}$.}
\label{Stacked_Unstacked}
\end{figure}

To investigate how the $\rm H_2$ content (and associated quantities) changes as a function of radius, it is important to recover CO emission in the outermost parts of galaxies, which  host the faintest CO emission. To achieve this, we perform a spectral stacking of the $^{12}$CO ($J=2-1$) emission line using the \hi\ velocities to coherently align the spectra while integrating in rings. The CO spectral stacking recovers CO flux over a broad range of galactocentric radii, thus allowing us to test how both $R_{\rm mol}$ and SFE$_{\rm mol}$ are affected by the environment from the inner- to the outermost parts of VERTICO galaxies. 

We perform the CO emission line stacking procedure following the methodology described by \cite{Villanueva2021} (see \S 3.1), which is based on the approach detailed in \cite{Schruba2011}. The method relies on using the \hi\, velocity data to define the velocity range for integrating CO emission. The key assumption of this method is that both the \hi\ and CO velocities are similar at any galaxy location, which is consistent with the results shown by \cite{Levy2018} for star-forming galaxies selected from the Extragalactic Database for Galaxy Evolution (EDGE) and the Calar Alto Legacy Integral Field Area (CALIFA) surveys \citep[The EDGE-CALIFA survey,][]{Bolatto2017}. Since this may not be the case for cluster galaxies due to environmental effects perturbing the \hi\, and H$_2$ in a distinct way, we test this by computing the differences between the CO(2--1) and \hi\, velocities, $\Delta V= V_{\rm CO(2-1)} - V_{\rm HI}$, in spaxels within the 38 VERTICO galaxies included in this work. Figure \ref{Vco_Vhi} shows that typically the differences between the atomic gas and the CO velocities are almost always smaller than the size of the velocity integration window (discussed below). To make sure that we recover as much CO intensity as possible, we implement a ``smart stacking''. We only take the stacked CO intensities in annuli where we have a strong \hi\, signal (i.e., \hi\, surface densities $>1$ M$_\odot$ pc$^{-2}$, yielding reliable \hi\, velocity measurements); otherwise, we take the unstacked CO intensities. Because of beam smearing in the \hi, intensities in the innermost parts of the galaxies ($r\lesssim0.3r_{25}$) can produce unreliable \hi\, velocity estimates, we employ stacked CO intensities only in annuli where their SNR is larger than that of the unstacked data.

Since we are interested in radial variations of the galaxy properties of VERTICO galaxies, we stack in radial bins $\sim0.1r_{25}$ wide. In practice, galactocentric radius is usually a well-determined observable and it is covariant with other useful local parameters, which makes it a very useful ordinate \citep{Schruba2011}. As discussed in \cite{Villanueva2021} and later in this section, after shifting the CO spectra to match the \hi\ velocity in each spaxel within a given annulus, we integrate over a spectral window designed to minimize missing CO flux to compute the CO (2--1) line emission intensity, $I_{\rm CO(2-1)}$. To define the integration window for the annuli, we use the third-order polynomial included in the top panel of Figure 2 in \cite{Villanueva2021}. They analyze the variation of molecular velocity dispersion as a function of radius for a sample of galaxies, and found a velocity envelope that is characterized as a polynomial. We use the same relation here to define our integration window.

Figure \ref{Stacked_Example} shows the usefulness of the stacking procedure in recovering the average CO(2--1) line emission. As an example, we show the average CO spectrum of NGC $4536$ within an annulus that spans from 0.6 to 0.7$r_{25}$. The left panel contains the average CO spectra within the given annulus using the observed velocity frame, while the right panel shows the average CO spectra after shifting by the observed \hi\ velocity. This procedure allows us to co--add CO intensities coherently and minimize noise. Figure \ref{Stacked_Example} also shows the best Gaussian fit for the averaged-stacked spectra (green-dashed line); as can be seen, the signal-to-noise ratio in the measurement of CO velocity-integrated intensity is lower and without performing the stacking procedure the CO line emission is not clearly detected. To quantify the improvement of the flux recovery, we compute the ratios between the final stacked and unstacked integrated CO(2--1) line intensity, $I_{\rm CO,Stack}/I_{\rm CO,Unstack}$, for each annulus. A histogram of the distribution of ratios is shown in Figure \ref{Stacked_Unstacked}. We find a median ratio of $\log_{10}[I_{\rm CO,Stack}/I_{\rm CO,Unstack}]\sim 0.21$; on average, the intensity in an annulus is increased by nearly $\sim 60$\%. We emphasize that most of the annuli with significant enhancements have weak CO emission and are found at large galactocentric radii. Thus, while stacking does not result in a large increase in total CO flux from a galaxy, it does extend the range of radii over which CO is detected, and results in a more accurate (larger) measurement of flux at larger radii. To compute the integrated flux uncertainties, we take the RMS from the emission-free part of the stacked CO spectra. We adopt a clipping level of 3 in order to consider a valid detection.  On average, we reach a characteristic rms noise of 0.1 mK at 10 km s$^{-1}$ (in the range of rms$\sim0.05-1$ mK), which corresponds to a sensitivity of $\sim$ 0.1 M$_{\odot}$ pc$^{-2}$ per 10 km s$^{-1}$ channel. Since we are interested in comparing of the radial profiles of the molecular gas, atomic gas, stellar mass, and star formation rate, it is appropriate to average over the entire annulus rather than limiting the average to spaxels where emission is detected.

\subsection{Basic Equations and Assumptions}
\label{Basic_equations}

The molecular gas surface density, $\Sigma_{\rm mol}$, is derived from the integrated CO intensity, $I_{\rm CO(2-1)}$, by adopting a constant CO-to-H$_2$ conversion factor, that is based on observations of the Milky Way; $X_{\rm CO} = 2 \times 10^{20}$ cm$^{-2}$ (K km s$^{-1}$)$^{-1}$, or $\alpha_{\rm CO,MW} = 4.3$ M$_\odot$ $(\rm K \, km \, s^{-1} \, pc^{2})^{-1}$ for the CO(J=1--0) line (\citealt{Walter2008}), following the analysis by \cite{Brown2021}. We also test how our results depend on the $\alpha_{\rm CO}$ prescription that we adopt by using the CO-to-H$_2$ conversion factor from Equation 31 in \citealt{Bolatto2013}:
\begin{equation}
\alpha_{\rm CO}  = 2.9 \exp \left ( \frac{+0.4}{Z' \Sigma^{100}_{\rm GMC}} \right ) \left ( \frac{\Sigma_{\rm total}}{100 \, \rm M_\odot \, pc^{2} } \right )^{-\gamma},
\label{eq:alpha_co}
\end{equation}

\noindent in M$_\odot$ $\rm (K \, km s^{-1} \, pc^{-2} )^{-1}$, $\gamma \approx 0.5$ for $\Sigma_{\rm total} > 100$ M$_\odot$ pc$^{-2}$ and $\gamma =0$ otherwise, and the metallicity normalized to the solar one, ${Z}' = \rm [O/H]/[O/H]_\odot$, where $\rm [O/H]_\odot = 4.9\times 10^{-4}$ \citep[][]{Baumgartner&Mushotzky2006}, $\Sigma^{100}_{\rm GMC}$ is the average surface density of molecular gas in units of 100 M$_\odot$ pc$^{-2}$, and $\Sigma_{\rm total}$ is the combined gas plus stellar surface density on kpc scales. Since we are interested in the global variations of the $\alpha_{\rm CO}({Z}',\Sigma_{\rm total})$, we mainly focus our analysis on its dependence on $\Sigma_{\rm total}$, since variations in this have the dominant effect for the regions studied in our sample. Therefore, we adopt a constant solar metallicity (i.e., ${Z}'=1.0$). 
We use the following expression to obtain $\Sigma_{\rm mol}$
\begin{equation}
\Sigma_{\rm mol}  = \frac{\alpha_{\rm CO}}{R_{21}} \cos (i)\, I_{\rm CO(2-1)},
\label{eq:mol}
\end{equation}

\noindent which adopts the average VERTICO survey's line luminosity ratio of $R_{21}=I_{\rm CO(2-1)}/I_{\rm CO(1-0)}=0.77\pm0.05$ obtained by \cite{Brown2021}, and $i$ is the inclination of the galaxy. This equation takes into account the mass ``correction'' due to the cosmic abundance of helium.

The atomic gas surface density, $\Sigma_{\rm atom}$, is computed from the integrated 21 cm line intensity taken from the VIVA survey \citep[][]{Chung2009}, $I_{\rm 21 cm}$, by using the following equation \citep[i.e., ][]{Leroy2008}

\begin{equation}
\frac{\Sigma_{\rm atom}}{{\rm M}_\odot \, {\rm pc}^{-2}} = 0.02\, \cos(i)\, \frac{I_{\rm 21cm}}{\rm K\, km\, s^{-1}},
\label{eq:atom}
\end{equation}

\noindent which includes both the inclination and a factor of 1.36 to account for the presence of helium.

Recent observational evidence has shown that molecular gas content and its distribution in the disk of cluster galaxies depends on the effect of the environment on the \hi\ distribution  \citep[e.g.,][]{Chung&Kim2014,Boselli2014a,Cortese2021,Zabel2022}. In order to characterize the behavior of the molecular gas as a function of the cluster environmental effects on the atomic gas, we use the \hi\ classification from \cite{Yoon2017} (hereafter \hi-Class). The classification is designed to quantify the perturbation level of atomic radial profiles based on morpho-kinematic \hi\ features and \hi-deficiency (e.g., \citealt{Haynes&Giovanelli1984,Cortese2021}) present in the VIVA survey. In total, 48 sources were selected by \cite{Yoon2017} to construct the classification, which are a good representation of Virgo galaxies undergoing various strengths of gas stripping. Note that this is different from classifying the galaxies in order of increasing \hi\ deficiency, although more highly disturbed galaxies tend to be more \hi\ deficient. According to \cite{Yoon2017}, the 38 VERTICO galaxies analyzed in this work can be categorized into the following five classes (including the number of galaxies in each of them):

\begin{figure*}
\hspace{-0.35cm}
  \includegraphics[width=18.2cm]{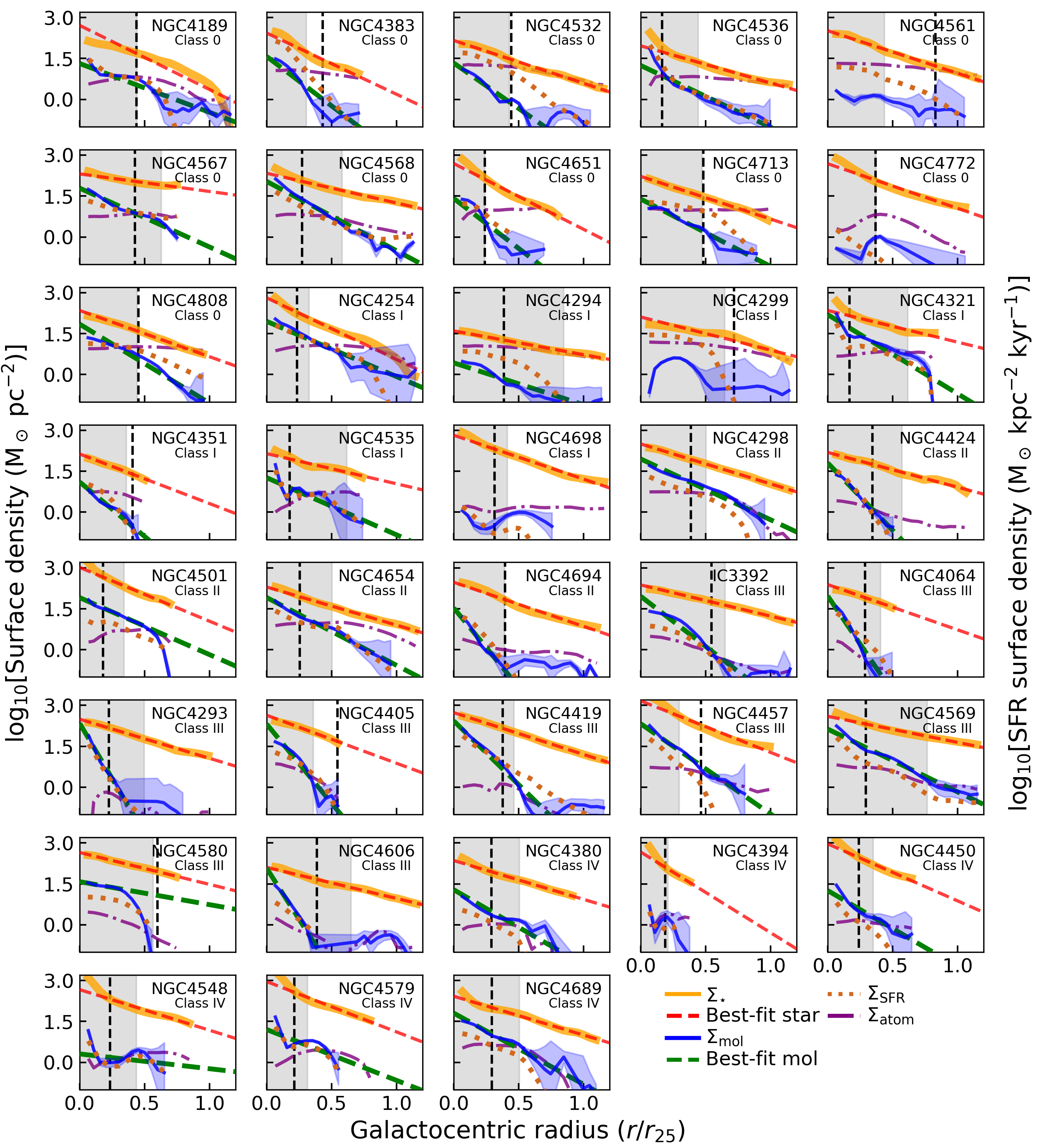}
  \caption{Stacked molecular gas ($\Sigma_{\rm mol}$, solid-blue line) and stellar ($\Sigma_{\star}$, solid-orange line) surface densities, in units of M$_\odot$ pc$^{-2}$, as a function of galactocentric radius, in units of $r_{25}$, for the 38 VERTICO galaxies analyzed in this work (sorted by \hi-Class). The shaded-blue area is the $\Sigma_{\rm mol}$ uncertainty. The brown-dotted line is the star formation rate surface density, $\Sigma_{\rm SFR}$. The purple-dashed line is the atomic gas surface density derived from \hi\ moment 0 maps at 15\arcsec \, resolution from the VIVA survey. The shaded-gray area is the region within the stellar effective radius $R_{\rm e,\star}$. The dashed-green and dashed-red lines represent the best-fit exponential profiles for $\Sigma_{\rm mol}$ and $\Sigma_{\star}$, respectively, when an exponential fit was appropriate. The vertical-dashed lines correspond to $r_{\rm gal}=3$kpc.}
 \label{Radial_profiles}
\end{figure*}

\begin{enumerate}\addtocounter{enumi}{-1}%[start=0]
    \item Class 0 (11 galaxies): The \hi \, profiles are symmetric, not truncated, and have extended and similar \hi content compared to most normal field galaxies. These are therefore the cases showing no definite signs of gas stripping due to the ICM.
\end{enumerate}

\begin{enumerate}{\Roman{enumi}}
    \item Class I (7 galaxies): One-sided \hi \, feature, such as a tail, and no truncation of the \hi \, disk within the relatively symmetric stellar disk; range of \hi \, deficiencies shown, but overall comparable to those of field galaxies.

   \item Class II (5 galaxies): A highly asymmetric \hi\ disk, with one-sided gas tails, extraplanar gas, and/or \hi\ disk truncation on at least one side of the stellar disk; quite deficient in \hi\, with an average of only $\sim$17\% of the typical \hi\ mass of a field counterpart.
    
   \item Class III (9 galaxies): A symmetric, but severely truncated \hi\ disk; extremely deficient in \hi\ with an average of $<$4\% of the \hi\ mass of a field galaxy counterpart.
    
    \item Class IV (6 galaxies): A symmetric \hi\ disk with marginal truncation within the radius of the stellar disk; lower \hi\ surface density than the other subclasses; quite deficient in \hi\, with on average $\sim$15\% of the \hi\ mass of a field galaxy counterpart.
\end{enumerate}

\noindent A more quantitative description of these galaxy categories can be found in \cite{Yoon2017}. Since a definition of the \hi-Classes based on a single criterion is not trivial, we also complement this classification by categorizing these five \hi-Classes in three broader groups as follows: i) Unperturbed galaxies (\hi-Class 0); ii) asymmetric galaxies (\hi-Classes I and II); and iii) symmetric-truncated galaxies (\hi-Classes III and IV). These \hi-Groups represent a powerful classification, which boost the statistics and provide a simpler analysis framework.

The spatially-resolved molecular-to-atomic gas ratio, $R_{\rm mol}$, is calculated as

\begin{equation}
R_{ \rm mol} = \frac{\Sigma_{\rm mol}}{\Sigma_{\rm atom}}. 
\label{eq:mol-to-atom}
\end{equation}

\noindent Similarly, we compute the spatially resolved molecular-to-stellar and the atomic-to-stellar ratios, $R^{\rm mol}_{\star}=\Sigma_{\rm mol}/\Sigma_{\star}$ and $R^{\rm atom}_{\star}=\Sigma_{\rm atom}/\Sigma_{\star}$, respectively, where $\Sigma_{\star}$ is the stellar surface density derived from the WISE band-1 data. To obtain the integrated values for these ratios (i.e. mass ratios), we integrate the surface densities for the molecular gas, atomic gas, and stars (assuming that they are distributed along a thin disk) to obtain the total masses as:

\begin{equation}
M_{i}(r<R_{\rm e,\star}) = 2 \pi \int^{R_{\rm e,\star}}_{0} \Sigma_{i}(r) \, r dr, 
\label{eq:global}
\end{equation}

\noindent where $R_{e,\star}$ is the effective radius for the stellar component (see \S \ref{Effective_radii} for more details), and $i=$ mol, atom, $\star$. We then calculate the integrated ratios as the ratio of the masses. We compute the resolved SFR surface density per unit molecular gas surface density, i.e. the star formation efficiency of the molecular gas (SFE$_{\rm mol}$, in units of yr$^{-1}$) for each annulus, 

\begin{equation}
{\rm SFE_{\rm mol}} = \frac{\Sigma_{\rm SFR}}{\Sigma_{\rm mol}},
\label{eq:sfe}
\end{equation}

\noindent where $\Sigma_{\rm SFR}$ is the resolved SFR surface density. We also obtain the integrated star formation efficiency of the molecular gas within $R_{\rm e,\star}$, SFE$_{\rm mol}(r<R_{\rm e,\star})$, using $\Sigma_{\rm SFR}(r)$, $\Sigma_{\rm mol}(r)$, and Equation \ref{eq:global}, so

\begin{equation}
{\rm SFE_{\rm mol}}(r<R_{\rm e}) = \frac{2 \pi \int^{R_{\rm e}}_{0} \Sigma_{\rm SFR}r dr}{M_{\rm mol}(r<R_{\rm e})}.
\label{eq:sfe_Re}
\end{equation}

\noindent Finally, to calculate the galactocentric radius of each annulus for each galaxy we use the inclinations from \cite{Brown2021} and we adopt the distance to the Virgo cluster of 16.5 Mpc from \cite{Mei2007}.

\subsection{CO Radial Profiles}
\label{CO_profiles}

As discussed in \S \ref{Stacking_CO}, we derive the molecular gas radial profiles by measuring the averaged azimuthal CO surface density, after performing a spectral stacking, in elliptical annuli  in the 9\arcsec \, CO(2--1) datacubes. Although CO radial profiles for VERTICO galaxies are already presented in \cite{Brown2021} and \cite{Zabel2022} (both using the same methodology described in \S4.3 of \citealt{Brown2021}), the CO spectral stacking expands the radial coverage and recovers faint CO emission especially in the outermost regions of most cases. Figure \ref{Radial_profiles} shows the molecular gas radial profiles derived in this work (blue-solid line), and their $\pm 1\sigma$ uncertainties (blue-shaded areas) for the 38 VERTICO galaxies selected here. Although radial profiles in \cite{Brown2021} and used in \cite{Zabel2022} are not corrected for inclination, they agree fairly with those included in Figure \ref{Radial_profiles} (particularly at $r<0.3 r_{25}$) when we multiply them by $\cos (i)$. Annuli are centered on the optical galaxy position and aligned with the major-axis position angle, taken from Table 1 in \cite{Brown2021}. After summing the velocity-integrated CO line emission pixel intensities in an annulus, we divide the sum by the total number of pixels to obtain the average $I_{\rm CO(2-1)}$ for the annulus. We then use Equation \ref{eq:mol} to obtain the molecular gas surface density, $\Sigma_{\rm mol}$. 

Galaxies in Figure \ref{Radial_profiles} are sorted by \hi-Class. Although $\Sigma_{\rm mol}$ tends to be lower than $\Sigma_{\rm atom}$ for \hi-Classes 0 and I (except for some galaxies with $\Sigma_{\rm mol}>\Sigma_{\rm atom}$ at $r\lesssim 0.3 r_{25}$), we note that the molecular gas seems to extend at least up to $\sim0.5 r_{\rm 25}$ (i.e., $\Sigma_{\rm mol}>1$ M$_{\odot}$ pc$^{-2}$; excluding NGC 4772, NGC 4299, and NGC 4698). Conversely, \hi-Classes II-IV show clear signs of truncation in both molecular and atomic gas radial profiles (except for NGC 4298, NGC 4654, and NGC 4569), with $\Sigma_{\rm mol}>\Sigma_{\rm atom}$ at any galactocentric radius. Interestingly, we note a systematic correlation in the truncation between $\Sigma_{\rm mol}$ and $\Sigma_{\rm atom}$ radial profiles with increasing \hi-Class.

\begin{figure*}
  \includegraphics[width=9.8cm]{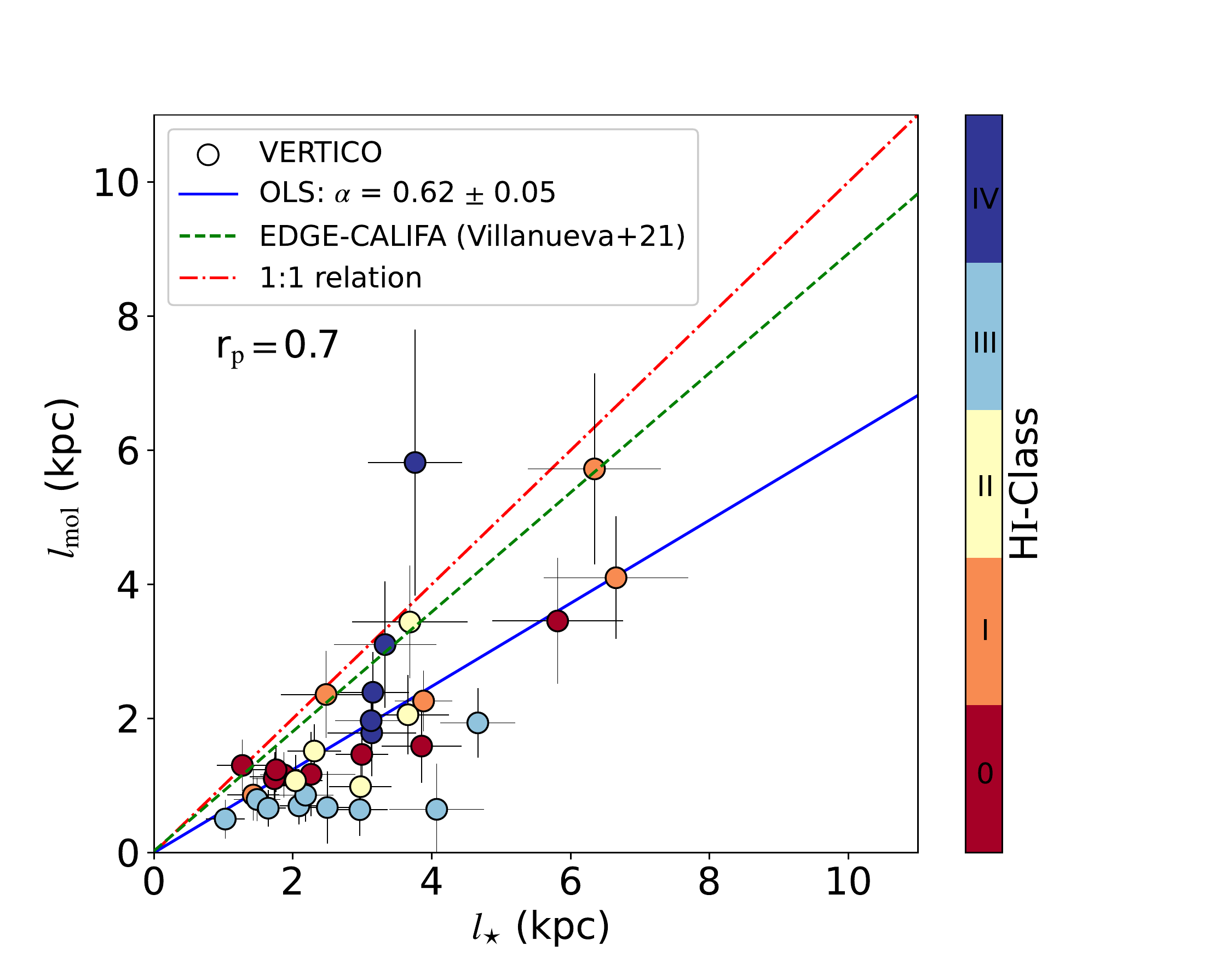}
  \includegraphics[width=9.8cm]{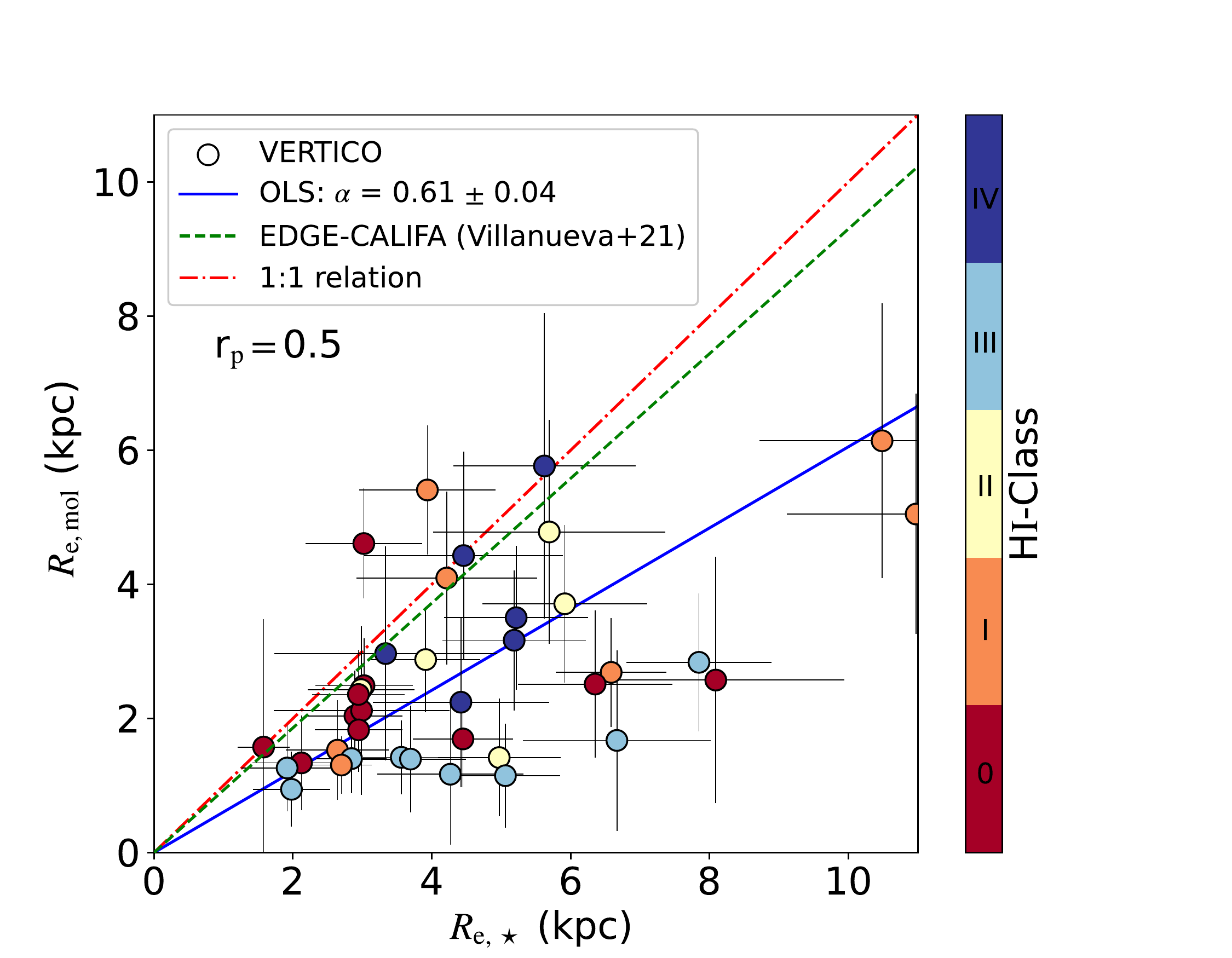}
  \caption{{\it Left:} Comparison between the stellar, ${\it l}_{\star}$, and molecular scale lengths, ${\it l}_{\rm mol}$, computed by fitting exponential profiles to the respective surface densities as a function of galactocentric radius. The colored circles correspond to 33 VERTICO galaxies with $\Sigma_{\rm mol}> 1$ M$_\odot$ pc$^{-2}$ for all the annuli within $0.25r_{25}$, color-coded by \hi-Class from \cite{Yoon2017}. The blue-solid line is the OLS linear bisector fit (forced through the origin) for them, and the dashed-dotted-red and dashed-green lines illustrate the 1:1 scaling and the OLS linear bisector fit for EDGE-CALIFA galaxies \citep[][]{Villanueva2021}, respectively. The `r$_{\rm p}$' value noted corresponds to the Pearson correlation parameter. {\it Right:} The molecular, $R_{\rm e,mol}$, vs stellar, $R_{\rm e,\star}$, effective radii, which enclose 50\% of the total molecular gas and stellar masses, respectively, for the 38 VERTICO galaxies analyzed in this work. Conventions are as in left panel.}
 \label{3_Exp_lenght}
\end{figure*}

To estimate how $\Sigma_{\rm mol}$ depends on the $\alpha_{\rm CO}$ prescription, we compute the ratio between the variable and the constant CO-to-H$_2$ conversion factors, $\alpha_{\rm CO}(\Sigma_{\rm total})/\alpha_{\rm CO,MW}$. We find that $\alpha_{\rm CO}(\Sigma_{\rm total})/\alpha_{\rm CO,MW}$ ranges from $ 0.2$ to $1.0$ in the region within $\sim0.6r_{25}$ (with a median of $0.95$); for $r>0.6r_{25}$, we find that $\alpha_{\rm CO}(\Sigma_{\rm total})/\alpha_{\rm CO,MW}=1.0$. Although these results show that $\Sigma_{\rm mol}$ depends on the adopted conversion factor, we note that $\alpha_{\rm CO}(\Sigma_{\rm total})$ has very small departures from $\alpha_{\rm CO,MW}$ for most of the annuli; consequently, the trends that we find in this work do not vary significantly due to the prescriptions for $\alpha_{\rm CO}$ selected for this work.  We emphasize however that our exploration of the effects of $\alpha_{\rm CO}$ dependence is limited and it deserves a careful analysis in future VERTICO projects.

Except for the spectral stacking used for CO, we implement the same method (averaging over all pixels in an annulus) for the star-formation rate, atomic gas, and stars. The latter two are shown in Figure \ref{Radial_profiles} by the purple-solid and orange-solid lines, respectively.

\section{Results and Discussion}
\label{S5_Results}

\subsection{Scale Lengths and Environment}
\label{Exponential_scale}

How do the relations between spatial distributions of the molecular gas and the stellar components depend on galaxy environment? Several studies have revealed the close relation between the spatial distribution of molecular gas and stars in galaxies selected from the field \citep[e.g.,][]{Young1995,Regan2001}. For instance, analyzing the molecular gas, $l_{\rm mol}$, and stellar, $l_{\star}$, exponential scale lengths (the brightness of the disc has fallen off by a factor of $e$, or $\sim2.71828$, from the center) for spiral galaxies selected from HERACLES \citep[][]{Leroy2009}, \cite{Leroy2008} showed a roughly 1:1 relation between them, with a best-fitting relation of $l_{\rm mol}=[0.9\pm0.2]\times l_{\star}$. A larger recent survey of 68 galaxies from the EDGE-CALIFA survey which covers a broad variety of morphologies finds a similar relation, $l_{\rm mol}=[0.89\pm0.04]\times l_{\star}$ \citep{Villanueva2021}.

\begin{table*}
\hspace{-1.cm}
\resizebox{0.95\linewidth}{!}{ 
\begin{tabular}{ccccccccc}
\hline\hline
Name & \hi-Class & $\log[M_{\rm mol}/M_{\odot}]$  & $\log[M_{\star}/M_{\odot}]$ & $l_{\rm mol}$ (kpc) & $l_{\star}$ (kpc) &  $R_{\rm e, mol}$ (kpc) & $R_{\rm e, \star}$ (kpc)\\
(1) &  (2)  & (3) & (4) & (5) & (6) & (7) & (8) \\
\hline
IC 3392 & III & 8.41$\pm$0.15 & 9.81$\pm$0.06 & 0.71$\pm$0.28 & 2.09$\pm$0.28 & 1.42 $\pm$0.55 & 3.56$\pm$0.55\\ 
 NGC 4064 & III & 8.44$\pm$0.19 & 10.01$\pm$0.09 & 0.67$\pm$0.54 & 2.50$\pm$0.53 & 1.17 $\pm$1.05 & 4.27$\pm$1.05\\ 
 NGC 4189 & 0 & 8.68$\pm$0.13 & 9.66$\pm$0.09 & 1.32$\pm$0.38 & 1.27$\pm$0.37 & 2.49 $\pm$0.71 & 3.02$\pm$0.73\\ 
 NGC 4254 & I & 9.87$\pm$0.10 & 10.41$\pm$0.08 & 2.36$\pm$0.65 & 2.48$\pm$0.65 & 4.09 $\pm$1.29 & 4.22$\pm$1.33\\ 
 NGC 4293 & III & 8.70$\pm$0.17 & 10.51$\pm$0.07 & 0.65$\pm$0.68 & 4.07$\pm$0.68 & 1.67 $\pm$1.35 & 6.67$\pm$1.36\\ 
 NGC 4294 & I & 7.91$\pm$0.07 & 9.57$\pm$0.05 & 2.26$\pm$0.45 & 3.88$\pm$0.41 & 2.69 $\pm$0.81 & 6.58$\pm$0.79\\ 
 NGC 4298 & II & 9.09$\pm$0.11 & 10.02$\pm$0.07 & 1.51$\pm$0.41 & 2.31$\pm$0.39 & 2.88 $\pm$0.78 & 3.91$\pm$0.78\\ 
 NGC 4299 & I & 7.53$\pm$0.17 & 9.27$\pm$0.08 & ...  & ... & 1.31 $\pm$0.43 & 2.7$\pm$0.44\\ 
 NGC 4321 & I & 9.85$\pm$0.11 & 10.82$\pm$0.05 & 4.10$\pm$0.91 & 6.65$\pm$1.04 & 5.05 $\pm$1.79 & 10.98$\pm$1.86\\ 
 NGC 4351 & I & 7.81$\pm$0.16 & 9.36$\pm$0.12 & 0.86$\pm$0.38 & 1.43$\pm$0.37 & 1.53 $\pm$0.74 & 2.64$\pm$0.74\\ 
 NGC 4380 & IV & 8.58$\pm$0.11 & 10.17$\pm$0.07 & 2.39$\pm$0.60 & 3.15$\pm$0.52 & 3.54 $\pm$1.08 & 5.22$\pm$1.04\\ 
 NGC 4383 & 0 & 8.42$\pm$0.14 & 9.58$\pm$0.07 & 1.13$\pm$0.37 & 1.74$\pm$0.36 & 1.34 $\pm$0.71 & 2.12$\pm$0.71\\ 
 NGC 4394 & IV & 7.81$\pm$0.23 & 10.32$\pm$0.14 & ...  & ... & 2.97 $\pm$1.62 & 3.34$\pm$1.60\\ 
 NGC 4405 & III & 8.30$\pm$0.17 & 9.62$\pm$0.11 & 0.51$\pm$0.29 & 1.03$\pm$0.28 & 0.94 $\pm$0.56 & 1.98$\pm$0.56\\ 
 NGC 4419 & III & 9.01$\pm$0.15 & 10.23$\pm$0.08 & 0.86$\pm$0.41 & 2.18$\pm$0.43 & 1.39 $\pm$0.79 & 3.69$\pm$0.83\\ 
 NGC 4424 & II & 8.36$\pm$0.15 & 9.93$\pm$0.06 & 0.98$\pm$0.45 & 2.97$\pm$0.45 & 1.42 $\pm$0.88 & 4.97$\pm$0.88\\ 
 NGC 4450 & IV & 8.65$\pm$0.11 & 10.69$\pm$0.07 & 1.79$\pm$0.64 & 3.13$\pm$0.64 & 2.24 $\pm$1.27 & 4.42$\pm$1.27\\ 
 NGC 4457 & III & 9.02$\pm$0.13 & 10.36$\pm$0.08 & 0.79$\pm$0.32 & 1.49$\pm$0.33 & 1.26 $\pm$0.65 & 1.92$\pm$0.65\\ 
 NGC 4501 & II & 9.69$\pm$0.12 & 10.99$\pm$0.08 & 3.44$\pm$0.84 & 3.68$\pm$0.83 & 4.78 $\pm$1.67 & 5.69$\pm$1.67\\ 
 NGC 4532 & 0 & 8.29$\pm$0.08 & 9.51$\pm$0.07 & 1.16$\pm$0.34 & 1.87$\pm$0.34 & 2.04 $\pm$0.68 & 2.89$\pm$0.68\\ 
 NGC 4535 & I & 9.46$\pm$0.13 & 10.58$\pm$0.06 & 5.72$\pm$1.42 & 6.34$\pm$0.96 & 6.14 $\pm$2.05 & 10.48$\pm$1.76\\ 
 NGC 4536 & 0 & 9.35$\pm$0.11 & 10.35$\pm$0.06 & 3.45$\pm$0.94 & 5.81$\pm$0.94 & 2.57 $\pm$1.84 & 8.09$\pm$1.85\\ 
 NGC 4548 & IV & 8.96$\pm$0.15 & 10.68$\pm$0.06 & 5.82$\pm$1.98 & 3.76$\pm$0.68 & 5.77 $\pm$2.28 & 5.62$\pm$1.31\\ 
 NGC 4561 & 0 & 7.31$\pm$0.13 & 9.33$\pm$0.07 & ...  & ... & 1.57 $\pm$1.91 & 1.58$\pm$0.37\\ 
 NGC 4567 & 0 & 8.74$\pm$0.12 & 10.13$\pm$0.06 & 1.46$\pm$0.38 & 2.99$\pm$0.38 & 1.72 $\pm$0.72 & 4.45$\pm$0.72\\ 
 NGC 4568 & 0 & 9.43$\pm$0.12 & 10.38$\pm$0.06 & 1.59$\pm$0.55 & 3.85$\pm$0.57 & 2.51 $\pm$1.10 & 6.35$\pm$1.11\\ 
 NGC 4569 & III & 9.53$\pm$0.09 & 10.74$\pm$0.05 & 1.93$\pm$0.52 & 4.66$\pm$0.54 & 2.84 $\pm$1.03 & 7.85$\pm$1.05\\ 
 NGC 4579 & IV & 9.31$\pm$0.14 & 10.89$\pm$0.07 & 3.12$\pm$0.94 & 3.33$\pm$0.74 & 4.43 $\pm$1.55 & 4.46$\pm$1.43\\ 
 NGC 4580 & III & 8.58$\pm$0.15 & 9.94$\pm$0.07 & 0.66$\pm$0.27 & 1.65$\pm$0.25 & 1.41$\pm$0.51 & 2.84$\pm$0.51\\ 
 NGC 4606 & III & 8.20$\pm$0.14 & 9.85$\pm$0.06 & 0.64$\pm$0.39 & 2.96$\pm$0.44 & 1.15 $\pm$0.78 & 5.06$\pm$0.79\\ 
 NGC 4651 & 0 & 8.77$\pm$0.16 & 10.27$\pm$0.11 & 1.17$\pm$0.63 & 2.26$\pm$0.64 & 2.12 $\pm$1.26 & 2.99$\pm$1.26\\ 
 NGC 4654 & II & 9.33$\pm$0.10 & 10.23$\pm$0.07 & 2.06$\pm$0.59 & 3.66$\pm$0.59 & 3.71 $\pm$1.18 & 5.92$\pm$1.19\\ 
 NGC 4689 & IV & 9.06$\pm$0.11 & 10.24$\pm$0.08 & 1.97$\pm$0.55 & 3.13$\pm$0.52 & 3.17 $\pm$1.05 & 5.19$\pm$1.03\\ 
 NGC 4694 & II & 8.29$\pm$0.07 & 9.92$\pm$0.07 & 1.07$\pm$0.38 & 2.04$\pm$0.39 & 2.43 $\pm$0.76 & 2.98$\pm$0.77\\ 
 NGC 4698 & I & 8.09$\pm$0.13 & 10.46$\pm$0.08 & ...  & ... & 5.41 $\pm$0.96 & 3.94$\pm$0.98\\ 
 NGC 4713 & 0 & 8.33$\pm$0.11 & 9.51$\pm$0.08 & 1.13$\pm$0.32 & 1.73$\pm$0.31 & 1.83 $\pm$0.63 & 2.95$\pm$0.63\\ 
 NGC 4772 & 0 & 7.85$\pm$0.14 & 10.19$\pm$0.07 & ...  & ... & 4.61 $\pm$0.82 & 3.01$\pm$0.84\\ 
 NGC 4808 & 0 & 8.74$\pm$0.11 & 9.61$\pm$0.08 & 1.24$\pm$0.33 & 1.76$\pm$0.33 & 2.36 $\pm$0.67 & 2.94$\pm$0.67\\   
\hline
\end{tabular}}
\caption{Main properties of the 38 VERTICO galaxies analyzed in this work. The columns are (1) galaxy name; (2) \hi-Class from \cite{Yoon2017}; (3) logarithm of the total molecular gas mass derived as explained in Section \ref{Exponential_scale}; (4) logarithm of the total stellar mass derived as explained in Section \ref{Effective_radii}; (5) exponential scale length of the molecular gas; (6) exponential scale length of the stars; (7) effective radius of the molecular gas; (8) effective radius of the stars.}
\label{Table}
\end{table*}

To characterize the distributions, we compute $l_{\rm mol}$ and $l_{\star}$ using our molecular gas and stellar radial profiles. From the 38 galaxies with $i<75^{\circ}$, we have selected VERTICO galaxies with $\Sigma_{\rm mol}$ and $\Sigma_{\star}$ radial profiles well described by an exponential profile and with $\Sigma_{\rm mol}> 1$ M$_\odot$ pc$^{-2}$ for all annuli within $0.25r_{25}$. These fits are shown by the green and red dashed lines in Figure \ref{Radial_profiles}. We have rejected annuli with $r_{\rm gal}<1.5$ kpc to avoid the central regions that may be susceptible to significant variations of $\alpha_{\rm CO}$ \citep[e.g.,][]{Sandstrom2013}, or to breaks in the exponential scale lengths (particularly for stars) due to bulges \citep{Regan2001}. We obtain $l_{\rm mol}$ and $l_{\star}$ for a subsample of 33 galaxies that fulfill the selection criteria mentioned above; the relation between them, colored by \hi-Class, is shown in the left panel of Figure \ref{3_Exp_lenght}.  We observe a fairly strong correlation between $l_{\rm mol}$ and $l_{\star}$ (Pearson r$_{\rm p}=0.7$; $p$-value$<0.01$). The left panel of Figure \ref{3_Exp_lenght} also contains the ordinary least-squares (OLS; blue solid line) bisector fit for $y=\alpha x$ weighted by the uncertainties for the $l_{\rm mol}$ and $l_{\star}$ points. Columns (5) and (6) of Table \ref{Table} correspond to the $l_{\rm mol}$ and $l_{\star}$ values, respectively, for the 33 galaxies included on this section.
We find that $l_{\rm mol}=(0.62\pm0.05)\times l_{\star}$ ($\sim$3:5 relation), much shallower than the almost 1:1 relation between $l_{\rm mol}$ and $l_{\star}$ for (mostly) field EDGE-CALIFA galaxies (green-dashed line in left panel of Fig. \ref{3_Exp_lenght}). When we use a variable prescription of $\alpha_{\rm CO}(\Sigma_{\rm total})$ (Equation \ref{eq:alpha_co}), we obtain $l_{\rm mol}=(0.66\pm0.05)\times l_{\star}$, which is in agreement with the fixed $\alpha_{\rm CO}$. We note that while $l_{\rm mol}$ values for \hi-Classes 0, I, and II tend to be similar to those for $l_{\star}$, they seem to concentrate significantly below the EDGE-CALIFA spirals trend for \hi-Classes III and IV. This implies that the high-density environment of the Virgo Cluster has a measurable effect in compacting the spatial extension of the molecular gas.

Our results are consistent with studies performed in Virgo galaxies by \cite{Boselli2014a}, who analyze the relation between the CO-to-stellar ($i$-band) isophotal diameter ratio, $\rm D({\rm CO})_{\rm iso}/D({\it i})$, and \hi\ deficiency in galaxies selected from the {\it{Herschel}} Reference Survey \citep[HRS;][]{Boselli2010}. They find a systematic decrease of $\rm D({\rm CO})_{\rm iso}/D({\it i})$ with increasing \hi\ deficiency, which suggests that environmental effects act on both the molecular gas and \hi\ simultaneously, particularly constraining the H$_2$ content to galaxy centers (see also Fig. 3 and Fig. 4 in \citealt{Zabel2022} for similar results). This may be attributed to the outside-in ram pressure detected previously in Virgo galaxies, which can compress the atomic gas and increase the molecular gas production \citep[e.g., NGC 4548, 4522, 4330;][]{Vollmer1999,Vollmer2008,Vollmer2012b}. This mechanism also could create a drag that causes gas to lose angular momentum and drift in. \cite{Mok2017} also find a similar result for galaxies selected from the James Clerk Maxwell Nearby Galaxies Legacy Survey (NGLS; \citealt{Wilson2012}). Using $^{12}$CO($J=3-2$) data, they find steeper H$_2$ radial profiles in Virgo galaxies than for their field counterparts.

\begin{figure*}
\includegraphics[width=18.cm]{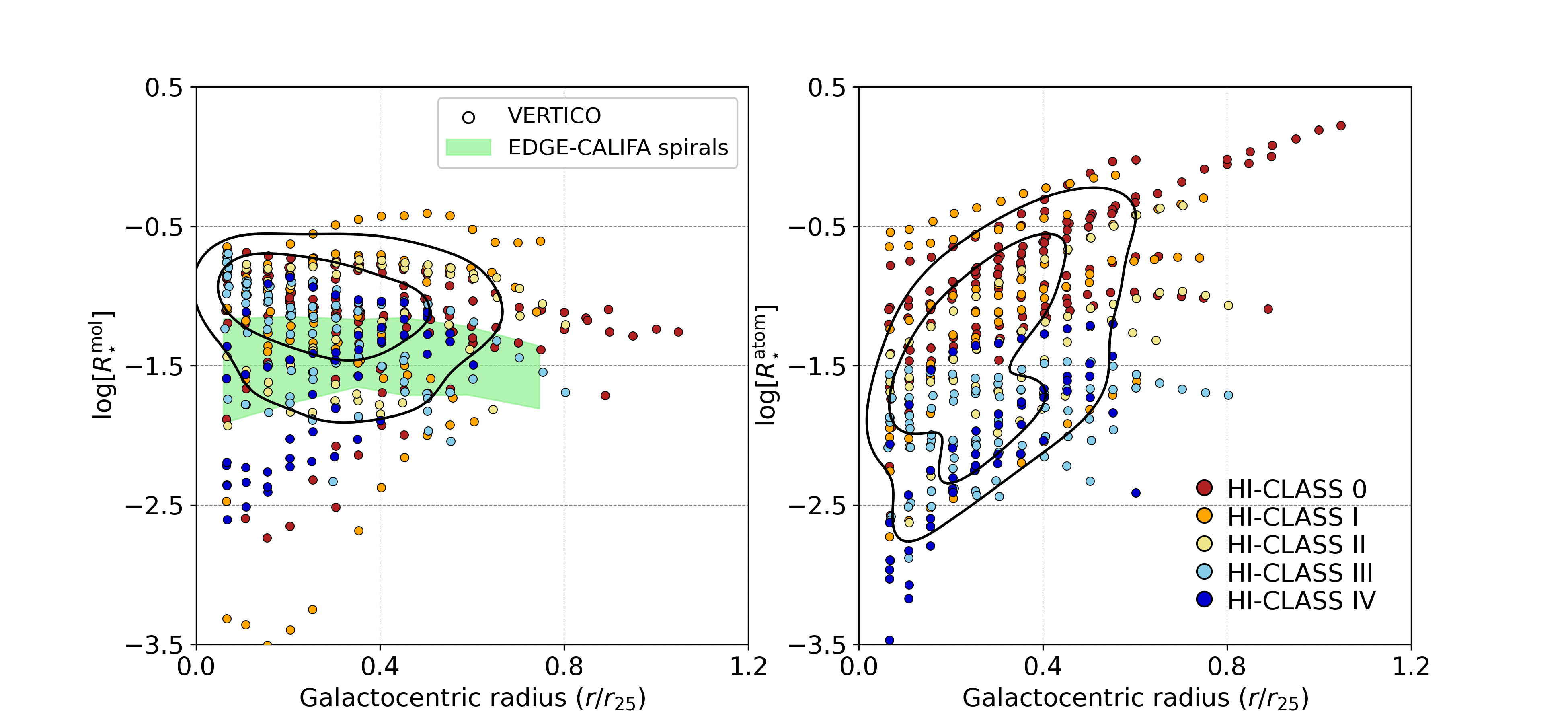}
\caption{{\it Left:} The resolved molecular-to-stellar mass ratio $R^{\rm mol}_{\star}$ colored by \hi-Class vs galactocentric radius for annuli within the 38 VERTICO galaxies analyzed in this work. The black contours  enclose, from outside-in, the 66\% and 33\% of the $R^{\rm mol}_{\star}$ of the points. The green shaded area is the range covered by EDGE-CALIFA spiral galaxies within $1\sigma$ scatter about the mean values for radial bins of $0.1r_{25}$ wide. {\it Right:} The resolved atomic-to-stellar mass ratio $R^{\rm atom}_{\star}$ vs galactocentric radius. Conventions are as in left panel.}
\label{Rmol_atom_star_vs_Rgal}
\end{figure*}

\subsection{Effective Radii and Environment}
\label{Effective_radii}

Comparison of scale lengths requires both the molecular and stellar radial distributions to be well described by an exponential. Non-parametric methods can help evaluate whether the assumption of an exponential disk may affect the conclusions above. The right panel of Figure \ref{3_Exp_lenght} shows the relation between the effective radius of the molecular gas, $R_{\rm e, mol}$, and the stars, $R_{\rm e, \star}$. These are the radii that enclose 50\% of the total molecular gas and stellar mass, respectively, for the 38 VERTICO galaxies analyzed in this work. We determine the total mass of each by integrating the $\Sigma_{\rm mol}$ and $\Sigma_{\star}$ radial profiles out to a distance of $\leq 1.2 r_{25}$. The relation (Pearson r$_{\rm p} = 0.5$; $p$-value$<0.01$) between $R_{\rm e, mol}$ and $R_{\rm e, \star}$ shows larger scatter than that between exponential scale lengths, but it nonetheless confirms the significant compactness of the molecular gas distribution compared to that of the stars. We also note that, in general, galaxies of higher \hi-Class (particularly in \hi-Class III) tend to have smaller $R_{\rm e, mol}$ and are more compact relative to their stellar distribution. 

In summary, the analysis of the $\Sigma_{\rm mol}$ radial profiles shows that VERTICO galaxies ($R_{\rm e, mol}=[0.61\pm0.04]\times R_{\rm e, \star}$) are approximately 30\% smaller in CO relative to their stellar distributions than EDGE-CALIFA galaxies ($R_{\rm e, mol}=[0.93\pm0.05]\times R_{\rm e, \star}$; \citealt{Villanueva2021}). When we use a variable $\alpha_{\rm CO} (\Sigma_{\rm total})$ (Equation \ref{eq:alpha_co}), we obtain $l_{\rm mol}=(0.66\pm0.04)\times l_{\star}$, which agrees fairly well with the previous result. Similar results are found by \cite{Zabel2022}, who show that VERTICO galaxies with larger \hi\, deficiencies (i.e., upper \hi-Classes) have steeper and less extended molecular gas radial profiles, suggesting that the processes behind the atomic gas removal are also producing more centrally concentrated molecular gas radial profiles.

Table \ref{Table} summarizes the properties of the 38 VERTICO galaxies selected for this work, together with the values for $R_{\rm e, mol}$, and $R_{\rm e, \star}$ (hereafter $R_{\rm e}$). In addition, columns (4) and (5) list $M_{\rm mol}$ (which are in good agreement with those included in \citealt{Brown2021}) and $M_{\star}$, calculated from the radial profiles.

\subsection{$R_{\rm mol}$, SFE, and Environment}

In this section we test how local and global physical parameters affect both the molecular-to-atomic gas ratio and the star formation efficiency of the molecular gas. 

\begin{figure*}
\includegraphics[width=18.cm]{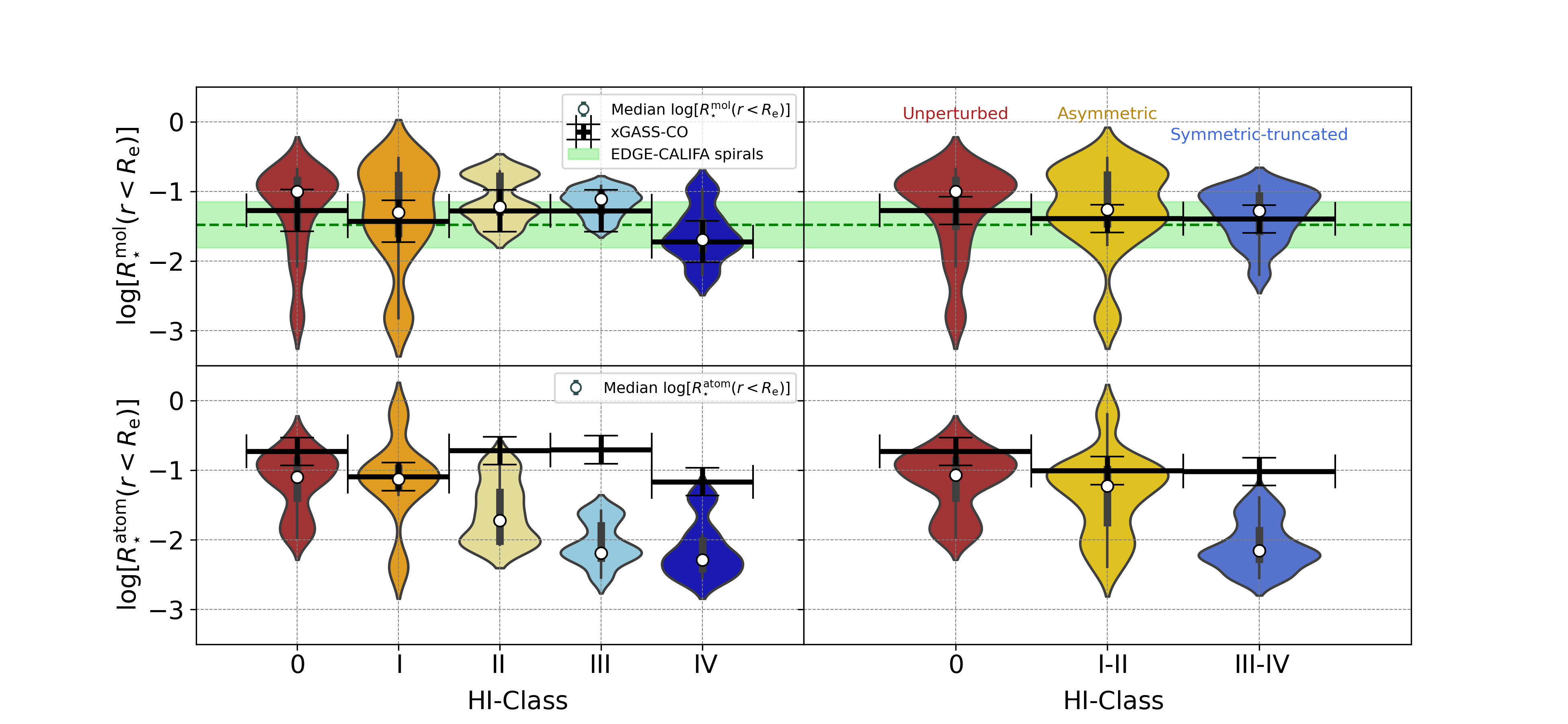}
\caption{{\it Left:} The molecular-to-stellar mass ratio within $R_{\rm e}$, $R^{\rm mol}_{\star}(r<R_{\rm e})=M_{\rm mol}(r<R_{\rm e})/M_{\star}(r<R_{\rm e})$ (top) and the atomic-to-stellar mass ratio within $R_{\rm e}$, $R^{\rm atom}_{\star}(r<R_{\rm e})=M_{\rm atom}(r<R_{\rm e})/M_{\star}(r<R_{\rm e})$ (bottom), vs \hi-Class defined by \cite{Yoon2017}. The black bars show the values obtained from the $M_{\rm mol}/M_{\star}$--$M_\star$ and $M_{\rm atom}/M_{\star}$--$M_\star$ relations for xGASS-CO MS galaxies from \cite{Saintonge&Catinella2022} using the mean stellar masses for the \cite{Yoon2017} \hi-Classes listed in Table \ref{Table_2}. The violin errorbars represent the distribution of values for each \hi-Class. The white dot is the median value of the distribution. The shaded-green area in the top panel is the $R^{\rm mol}_{\star}(r<R_{\rm e})$ range covered by EDGE-CALIFA spiral galaxies within $1\sigma$ scatter. {\it Right:} $R^{\rm mol}_{\star}(r<R_{\rm e}))$ (top) and the $R^{\rm atom}_{\star}(r<R_{\rm e})$ (bottom) vs \hi-Class after clustering them in three broader groups: i) unperturbed galaxies (\hi-Class 0), ii) asymmetric galaxies (\hi-Classes I and II galaxies); and iii) symmetric-truncated galaxies (\hi-Classes III and IV galaxies). Conventions are as in the left panel. While $R^{\rm mol}_{\star}(r<R_{\rm e})$ values for VERTICO galaxies are within the ranges covered by the xGASS-CO MS galaxies, $R^{\rm atom}_{\star}(r<R_{\rm e})$ values show a systematic decrease with increasing \hi-Class.}
\label{Rmol_atom_star_vs_HIClass}
\end{figure*}

To understand the effect of the environmental processes of the cluster on gas and star formation properties for VERTICO, we need a comparison sample, that represents galaxies in low-density environments. We compare with two such samples: 1) 64 galaxies selected from spatially resolved surveys of spiral galaxies with $\log[M_{\star}/$M$_{\odot}] = 9.1$--$11.5$ and morphologies spanning from Sa to Scd, EDGE-CALIFA survey; and 2) xGASS/xCOLD GASS \citep[hereafter xGASS-CO; ][]{Saintonge2017,Catinella2018}. For xGASS-CO we use the relations obtained in the analysis by \cite{Saintonge&Catinella2022} for main sequence (MS) galaxies. Each of these comparison samples have limitations that need to be kept in mind. The EDGE-CALIFA selected galaxies are mostly far-IR detected and rich in molecular gas (hence actively star-forming), and only a handful of them have resolved \hi\ observations. The xGASS-CO sample is the largest galaxy survey, but it is spatially unresolved.

We correct our calculations by the inclination of the galaxy (using a $\cos (i)$ factor) to represent physical ``face-on'' deprojected surface densities (see \S\ref{Basic_equations}). The EDGE-CALIFA spiral galaxies included here were selected to have $i<75^\circ$ \citep{Villanueva2021}.

\subsubsection{$R_{\rm mol}$ vs Radius and Environment}  
\label{Rmol_radius}

The left and right panels of Figure \ref{Rmol_atom_star_vs_Rgal} show the spatially resolved molecular-to-stellar $R^{\rm mol}_{\star}=\Sigma_{\rm mol}/\Sigma_{\star}$ and atomic-to-stellar $R^{\rm atom}_{\star}=\Sigma_{\rm atom}/\Sigma_{\star}$ ratios, in logarithmic space, as a function of galactocentric radius and colored by \hi-Class. Looking at the black contours, which enclose 66\% and 33\% of the points, we note that $R^{\rm mol}_{\star}$ follows a constant trend with radius and it has similar values to those covered by EDGE-CALIFA spirals. We also note that most of the galaxies show a systematic inside-out increase in $R^{\rm atom}_{\star}$ (Pearson r$_{\rm p}=0.53$ when considering all the points). 

The top and bottom panels of Figure \ref{Rmol_atom_star_vs_HIClass} show the molecular- and atomic-to-stellar mass ratios integrated out to $R_{\rm e}$, $R^{\rm mol}_{\star}(r<R_{\rm e})=M_{\rm mol}(r<R_{\rm e})/M_{\star}(r<R_{\rm e})$ and $R^{\rm atom}_{\star}(r<R_{\rm e})=M_{\rm atom}(r<R_{\rm e})/M_{\star}(r<R_{\rm e})$, respectively, as a function of \hi-Class. The upper panels show that the median $R^{\rm mol}_{\star}(r<R_{\rm e})$ values remain almost constant with \hi-Class but with a possible small decrease for \hi-Class IV galaxies. These values are in good agreement with those expected from the $M_{\rm mol}/M_{\star}$--$M_{\star}$ relation for xGASS-CO galaxies (black crosses; \citealt{Saintonge&Catinella2022}), and with EDGE-CALIFA spirals (green-shaded area). Interestingly, $R^{\rm atom}_{\star}(r<R_{\rm e})$ (bottom plots in Fig. \ref{Rmol_atom_star_vs_HIClass}) shows a systematic decrease from lower to upper \hi-Classes. While \hi-Classes 0 and I have atomic-to-stellar mass ratios with median values similar to that of xGASS-CO, \hi-Classes II, III, and IV have significantly lower $R^{\rm atom}_{\star}(r<R_{\rm e})$ values. 

We test how our results change when adopting a variable $\alpha_{\rm CO}(\Sigma_{\rm total})$. Although $R^{\rm mol}_{\star}$ values decrease $\sim0.2-0.3$ dex when using $\alpha_{\rm CO}(\Sigma_{\rm total})$ (particularly at $r\lesssim 0.6 r_{25}$), we note that the $R^{\rm mol}_{\star}$ trend does not vary significantly compared to that for $\alpha_{\rm CO,MW}$; similarly, $R^{\rm mol}_{\star}(r<R_{\rm e})$ values are still within the ranges covered by xGASS-CO when using the two $\alpha_{\rm CO}$ prescriptions. Despite there being a clear deficit in the integrated H\,{\sc i} content of the H\,{\sc i}-perturbed galaxies, our VERTICO data reveal that these galaxies do \emph{not} exhibit an appreciable deficit in H$_2$ mass.

\begin{figure*}
\hspace{-1.85cm}
\includegraphics[width=21.5cm]{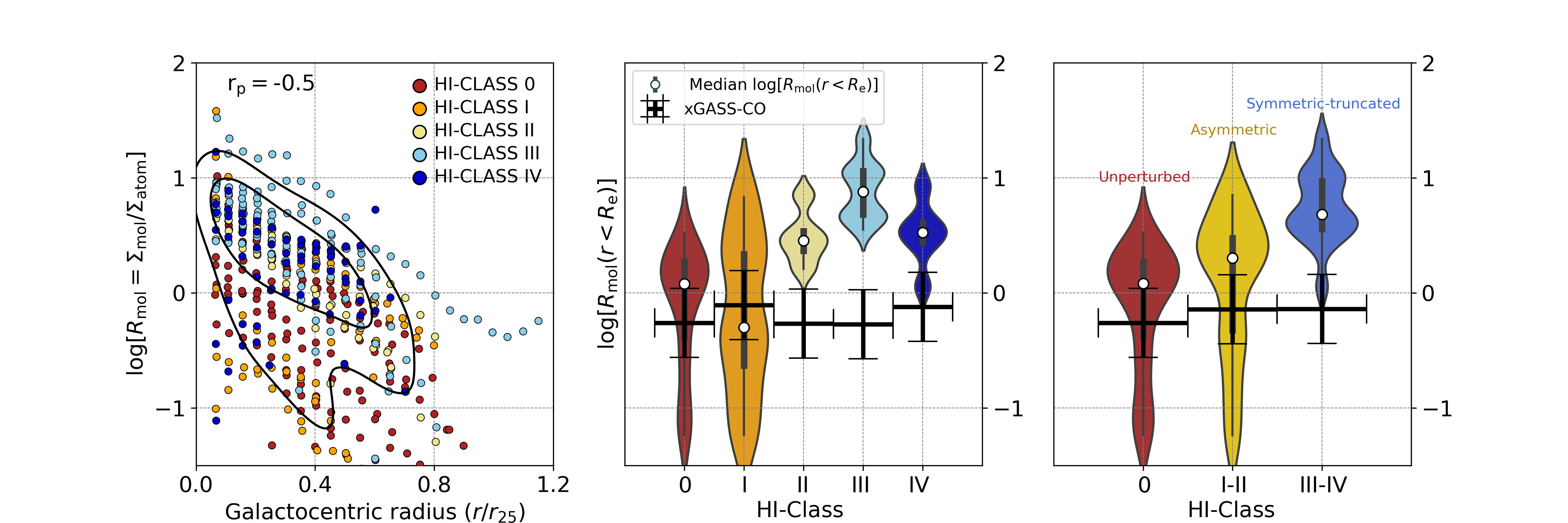}
\caption{{\it Left:} The resolved molecular-to-atomic gas ratio $R_{\rm mol}=\Sigma_{\rm mol}/\Sigma_{\rm atom}$ vs galactocentric radius, with circles colored by \hi-Class. The black contours enclose, from outside-in, the 66\% and 33\% of the points. {\it Middle:} The molecular-to-atomic gas ratio, in logarithmic space, computed using the molecular and atomic gas masses within $R_{\rm e}$, $R_{\rm mol}(r<R_{\rm e})=M_{\rm mol}(r<R_{\rm e})/M_{\rm atom}(r<R_{\rm e})$, vs \hi-Class. The black bars show the $R_{\rm mol}$ values obtained from the $M_{\rm mol}/M_{\rm atom}$--$M_\star$ relation for xGASS-CO MS galaxies from \cite{Saintonge&Catinella2022} using the mean stellar masses for the \cite{Yoon2017} \hi-Classes listed in Table \ref{Table_2}. There is an increase of $R_{\rm mol}(r<R_{\rm e})$ that is up to about an order of magnitude going from lower to higher \hi-Classes (e.g., from Class I to III); the more disturbed the \hi\, the higher the molecular-to-atomic ratio within $R_{\rm e}$. {\it Right:} $R_{\rm mol}(r<R_{\rm e})$ vs \hi-Class groups as in the right panel of Fig. \ref{Rmol_atom_star_vs_HIClass}. Conventions and symbols are as in Fig. \ref{Rmol_atom_star_vs_HIClass}.}
\label{Rmol_vs_Rgal}
\end{figure*}

The left panel of Figure \ref{Rmol_vs_Rgal} shows the spatially resolved molecular-to-atomic ratio $R_{\rm mol}=\Sigma_{\rm mol}/\Sigma_{\rm atom}$, in logarithmic space, as a function of galactocentric radius and colored by \hi-Class. To compute $R_{\rm mol}$, we use the $\Sigma_{\rm mol}$ radial profiles derived from the 15\arcsec\, CO(2--1) datacubes to match VIVA's \hi\, angular resolution. In general, the figure shows a decreasing trend for $R_{\rm mol}$ with radius (Pearson r$_{\rm p}=-0.5$; $p$-value$<0.01$). We note that $R_{\rm mol}$ values for \hi-Classes II, III, and IV (yellow, light blue, and blue solid dots, respectively) are on average higher than for \hi-Classes 0 and I (red and orange solid dots, respectively). Looking at Figure \ref{Radial_profiles}, we note that while lower \hi-Classes have similar $\Sigma_{\rm atom}$ to what is expected for normal field galaxies (e.g., $\Sigma_{\rm atom}\approx 6$ M$_\odot$ pc$^{-2}$; \citealt{Leroy2008}), upper \hi-Classes show notably lower atomic surface densities. Since \hi-Classes II, III, and IV are \hi\, deficient (as mentioned previously for $R_{\rm atom,\star}$), the enhancement of $R_{\rm mol}$ (at least within $R_{\rm e}$) appears to be due mainly to their poor atomic gas content. The significant scatter in $R_{\rm mol}$, particularly at $r_{\rm gal} \lesssim 0.3 r_{25}$, may be due to the strong environmental effects experienced by some VERTICO galaxies. We also compute the molecular-to-atomic gas mass ratio within $R_{\rm e}$, $R_{\rm mol}(r<R_{\rm e})=M_{\rm mol}(r<R_{\rm e})/M_{\rm atom}(r<R_{\rm e})$. The right panel of Figure \ref{Rmol_vs_Rgal} shows the relation between $R_{\rm mol}(r<R_{\rm e})$ (in logarithmic space) and \hi-Class. Although with a dip in \hi-Classes II and IV galaxies, there is a large systematic increase of $R_{\rm mol}(r<R_{\rm e})$ from lower to upper \hi-Classes which becomes even more clear for the broader \hi\ groups in the left panel of Fig. \ref{Rmol_vs_Rgal}. As noted previously for $R^{\rm mol}_{\star}$, $R_{\rm mol}$ values decrease $\sim0.2-0.3$ dex with $\alpha_{\rm CO}(\Sigma_{\rm total})$ at $r\lesssim 0.6 r_{25}$. Similarly, $R_{\rm mol}(r<R_{\rm e})$ values also decrease when using the variable $\alpha_{\rm CO}$ prescription. However, we still observe a clear systematic decrease of $R_{\rm mol}$ with radius, and $R_{\rm mol}(r<R_{\rm e})$ values are still within the ranges covered by xGASS-CO MS galaxies. This is consistent with the definition of \hi-Classes by \cite{Yoon2017} for VIVA galaxies, and it confirms that the increase in ratios is a result of the deficiency in \hi.  

These results can be summarized as follows:

\begin{enumerate}

    \item For unperturbed galaxies and with mild sign of \hi- perturbation (e.g., \hi-Classes 0 and I), $R^{\rm mol}_{\star}(r<R_{\rm e})$ and $R^{\rm atom}_{\star}(r<R_{\rm e})$ values are similar to those for xGASS-CO MS galaxies with similar stellar mass, and also have $R_{\rm mol}(r<R_{\rm e})$ values comparable with the latter. 
    
    \item Asymmetric galaxies in \hi\ or partially symmetric-truncated (\hi-Classes II and III) have $R^{\rm mol}_{\star}(r<R_{\rm e})$ values within the range covered by EDGE-CALIFA spirals and those expected from xGASS-CO relations. However, we note that $R^{\rm atom}_{\star}(r<R_{\rm e})$ values are up to $1.5$ dex lower than those for xGASS-CO MS galaxies. \hi-Class II and III galaxies also have $R_{\rm mol}(r<R_{\rm e})$ values significantly higher than those in stellar-mass matched xGASS-CO MS galaxies.
    
    \item \hi-symmetric-truncated galaxies (i.e., \hi-Class IV) galaxies show a possible decrease in $R^{\rm mol}_{\star}(r<R_{\rm e})$ compared with EDGE-CALIFA spirals, although still in good agreement with xGASS-CO MS galaxies. Similar to \hi-Classes II and III, \hi-Class IV has $R^{\rm atom}_{\star}(r<R_{\rm e})$ values drastically lower than the latter ($\sim1.0$ dex lower). \hi-Class IV galaxies also show an increase in $R_{\rm mol}(r<R_{\rm e})$ values compared to lower \hi-Classes (although a small decrease compared to \hi-Class III) and to those expected for xGASS-CO MS galaxies.
    
\end{enumerate}

\begin{table*}
\hspace{-2.cm}
\resizebox{1.1\linewidth}{!}{ 
\begin{tabular}{ccccccccc}
\hline\hline
Galaxy Class & log[$M_\star/M_\odot$] & log[$M_{\rm mol}/M_\odot$] & log[$M_{\rm atom}/M_\odot$] & log[SFR$/(M_\odot \rm\, yr^{-1})$] & log[$M^{\rm xGC}_{\rm mol,MS}/M_\odot$] & log[$M^{\rm xGC}_{\rm atom,MS}/M_\odot$] & log[$M_{\rm mol}/M_{\star}$]$_{\rm xGC}$ & log[$M_{\rm atom}/M_{\star}$]$_{\rm xGC}$\\
(1) & (2) & (3) & (4) & (5) & (6) & (7) & (8) & (9)\\
\hline
Class 0 & 9.66$\pm$0.38 & 8.59$\pm$0.56 & 9.29$\pm$0.29 & 0.58$\pm$0.72 & 8.46$\pm$0.22 & 9.31$\pm$0.46 & -1.28 & -0.54\\ 
 Class I & 10.41$\pm$0.61 & 8.07$\pm$0.92 & 9.21$\pm$0.36 & 0.32$\pm$1.61 & 9.01$\pm$0.27 & 9.55$\pm$0.39 & -1.61 & -1.14\\ 
 Class II & 10.02$\pm$0.43 & 9.04$\pm$0.51 & 8.69$\pm$0.46 & 0.13$\pm$0.21 & 8.79$\pm$0.20 & 9.38$\pm$0.48 & -1.27 & -0.73\\ 
 Class III & 10.01$\pm$0.34 & 8.42$\pm$0.35 & 7.63$\pm$0.44 & 0.27$\pm$0.47 & 8.78$\pm$0.23 & 9.38$\pm$0.47 & -1.27 & -0.72\\ 
 Class IV & 10.5$\pm$0.27 & 8.73$\pm$0.41 & 8.66$\pm$0.24 & 0.24$\pm$0.54 & 9.13$\pm$0.26 & 9.61$\pm$0.39 & -1.78 & -1.18\\ 
 EDGE-CALIFA & 10.57$\pm$0.45 & 9.45$\pm$0.47 & ... & ... & 9.16$\pm$0.26 & 9.64$\pm$0.39 & -1.89 & -1.21\\  
\hline
\end{tabular}}
\caption{Median and 1$\sigma$ scatter values of the mass and star formation rate distributions for the galaxy groups in column (1). The columns are: (2) logarithm of the median stellar mass; (3) logarithm of median molecular gas mass; (4) logarithm of median atomic gas mass; from Class 0 to Class IV, atomic gas masses are taken from \cite{Brown2021}; (5) logarithm of total median star formation rate; (6) logarithm of total molecular gas mass derived from the $M_{\rm mol}$--$M_{\star}$ relation for main sequence xGASS-CO galaxies by \cite{Saintonge&Catinella2022}, using the stellar mass from column (2); (7) logarithm of total atomic gas mass derived similarly as in column (6); (8) logarithm of the total molecular-to-stellar ratio derived from the $M_{\rm mol}/M_{\star}$--$M_{\star}$ relation for xGASS-CO galaxies by \cite{Saintonge&Catinella2022}, using the stellar mass from column (2), as shown by the black errorbars in the top panel of the right plot in Fig. \ref{Rmol_atom_star_vs_Rgal}; (9) logarithm of the total atomic-to-stellar ratio derived similarly as in column (8). For both columns (8) and (9), according to \cite{Saintonge&Catinella2022} the scatter is within $\sim0.3$ dex.}
\label{Table_2}
\end{table*}

These results suggest that even though environmental processes act on both the molecular and the atomic gas (at least within $R_{\rm e}$), the latter is affected in a different manner than the former. Table \ref{Table_2}, which includes a compilation of masses for all \hi-Classes, indicates that VERTICO galaxies have lower $M_{\rm mol}$ compared to EDGE-CALIFA spirals (although the latter is slightly biased towards molecule-rich galaxies). VERTICO galaxies also seem to have lower $M_{\rm mol}$ values than those expected from the $M_{\rm mol}-M_{\star}$ relation for main sequence xGASS-CO galaxies, although without significant variations with \hi-Class or $M_{\star}$. We note that $M_{\rm atom}$ values for \hi-Classes II, III, and IV are notably lower than for xGASS-CO. Similar results are found by \cite{Zabel2022}, who do not observe a statistically significant correlation between H$_2$ and \hi\, deficiencies. They also note that VERTICO galaxies tend to be H$_2$ deficient when compared to main sequence galaxies from xGASS-CO. These results suggest that even though environmental processes affect both the molecular and the atomic gas simultaneously, the level of \hi\, perturbation in VERTICO galaxies does not necessarily modulate the molecular gas content. The large change in $R_{\rm mol}(r<R_{\rm e})$ with environment (e.g., a factor of $\sim10$ between \hi-Classes 0 and III) also indicates that galaxy evolution simulations should factor this in if they want to trace \hi/H$_2$ phases in dense environments. Also, while $R^{\rm atom}_{\star}$ values for \hi-Class IV galaxies are on average the lowest, particularly at $r\lesssim 0.4r_{25}$, their $R^{\rm mol}_{\star}$ values show a significant decrease towards the centers. These results suggest that \hi-Class IV galaxies could be tracing the population where the environment is starting to impact significantly the molecular gas content.

Although several environmental mechanisms could potentially explain the results previously shown, the most likely mechanism is ram pressure (at least in \hi-Classes I-III). Out of the 38 VERTICO galaxies selected for this work, at least five of them are well studied cases of RPS: NGC 4501, NGC 4548, NGC 4569, NGC 4579, and NGC 4654 \citep[e.g.,][ see also \citealt{Boselli2022} and references therein]{Cayatte1994,Boselli2006,Vollmer2012a,Boselli2016,Lizee2021}. In particular, \cite{Mok2017} report significantly higher $R_{\rm mol}$ values for Virgo galaxies than for field galaxies. They attribute this to environmental processes, either by inward flows of molecular gas, H$_2$ not being as efficiently stripped as the atomic gas, and/or \hi\, migration to the galaxy center where it can be more easily converted into H$_2$. This is also supported by \cite{Moretti2020}, who analyze four jellyfish galaxies from the GAs Stripping Phenomena survey with MUSE (GASP; \citealt{Poggianti2017}). They propose that gas compression caused by ram pressure can trigger the conversion of large amounts of \hi\ into the molecular phase in the disk, which may imply that \hi\ is just partially stripped. The results found by \cite{Zabel2022} also support this idea, and suggest that RPS in VERTICO galaxies could potentially drive outside-in gas migration and may contribute to produce steeper H$_2$ radial profiles, as seen in \cite{Mok2017}. However, thermal evaporation (\citealt{Cowie&Songaila1977}; see also \citealt{Cortese2021} for a detailed description) or starvation have also been proposed to explain the effects of the environment in galaxies with symmetric-truncated \hi\, radial profiles (e.g., \hi-Class IV galaxies; see also \S\,5.1 in \citealt{Zabel2022}). Observational and theoretical evidence has shown that thermal evaporation can complement the viscous stripping in removing the cold gas from the disk \citep[e.g.,][]{Bureau&Carignan2002,Roediger&Hensler2005,Boselli&Gavazzi2006,Randall2008}. Consequently, it can affect the entire gas disk (all at the same time), leading to marginally truncated (but symmetric) gas disks with low surface density. Thermal evaporation is, therefore, a good candidate to explain the offset in $R_{\rm mol}(r<R_{\rm e})$ for \hi-Class IV galaxies (middle panel of Fig. \ref{Rmol_vs_Rgal}) when compared to lower \hi-Classes.

\begin{figure}
\hspace{-.3cm}
\includegraphics[width=9.5cm]{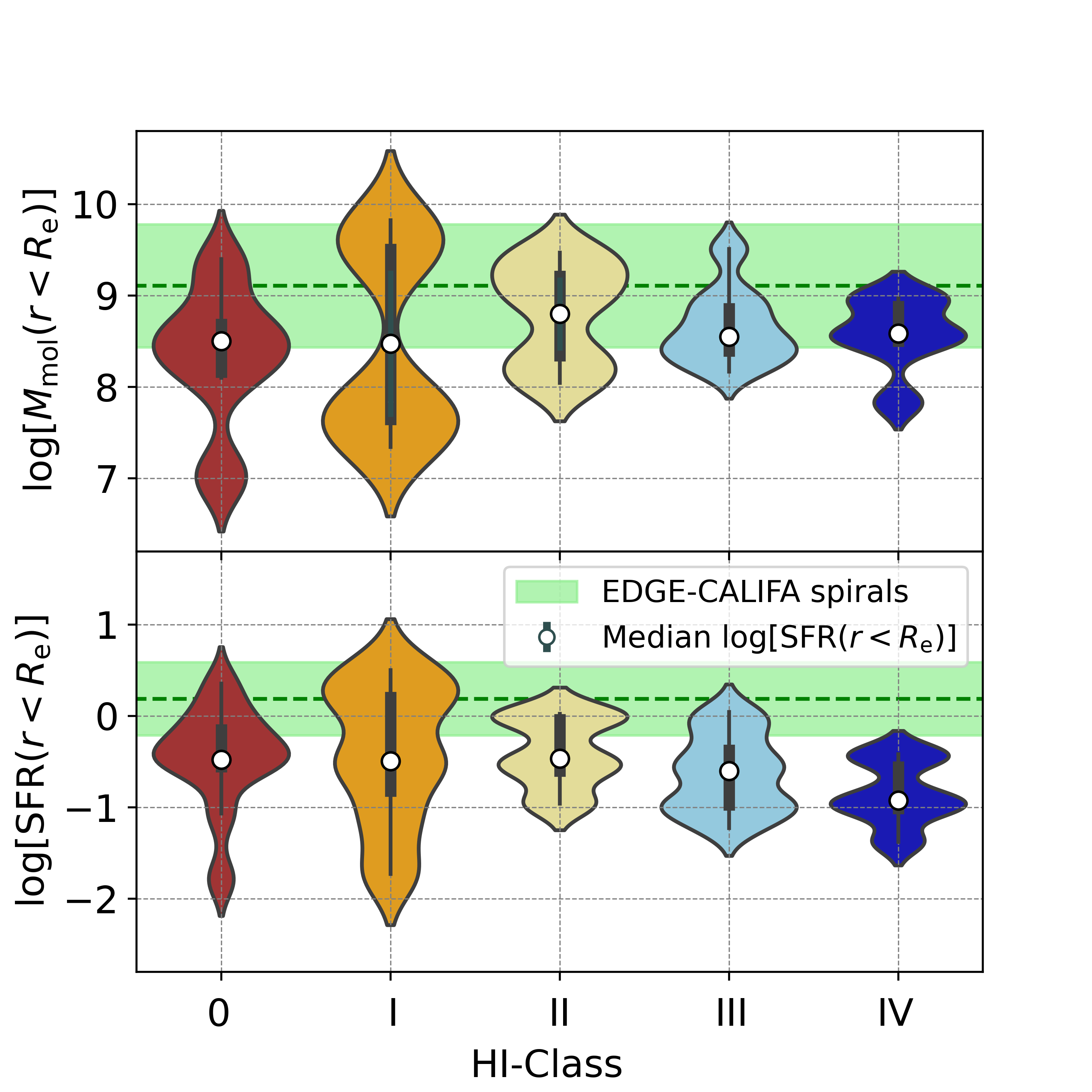}
\caption{{\it Top:} The molecular gas mass within $R_{\rm e}$, $M_{\rm mol}(r<R_{\rm e})$ (M$_\odot$), vs \hi-Class. {\it Bottom:} The star formation rate within $R_{\rm e}$, SFR$(r<R_{\rm e})$ (M$_\odot$ yr$^{-1}$), vs \hi-Class. Conventions are as in left panel of Fig. \ref{Rmol_atom_star_vs_HIClass}. While $M_{\rm mol}(r<R_{\rm e})$ remains almost constant, there is a systematic decrease of SFR$(r<R_{\rm e})$ with \hi-Class (particularly from \hi-Classes II to IV).}
\label{SFEmol_vs_Rgal_2}
\end{figure}

\begin{figure*}
\hspace{-1.85cm}
\includegraphics[width=21.cm]{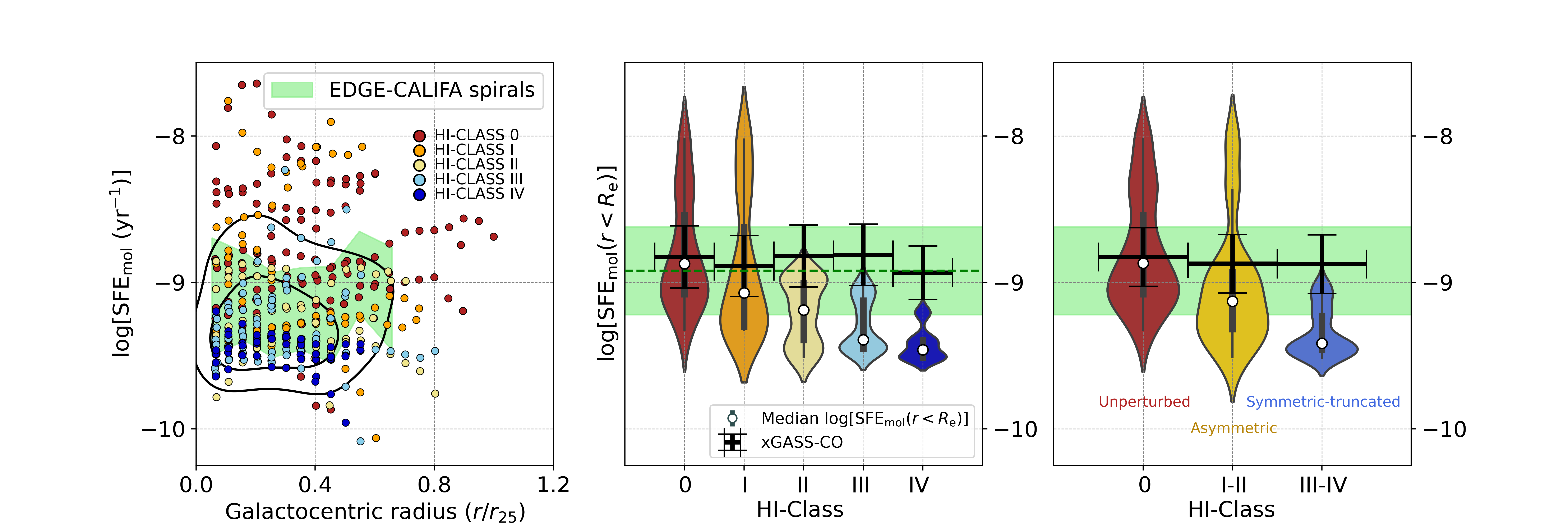}
\caption{{\it Left:} The resolved star formation efficiency of the molecular gas, SFE$_{\rm mol}$, vs galactocentric radius. The black contours enclose, from outside-in, the 66\% and 33\% of the points. {\it Middle:} The star formation efficiency of the molecular gas within $R_{\rm e}$, ${\rm SFE}_{\rm mol}(r<R_{\rm e})$, vs \hi-Class. The horizontal black bars are the ${\rm SFR}/M_{\rm mol}$ median values for \hi-Classes 0, I, II, III, and IV VERTICO galaxies using the ${\rm SFR}/M_{\rm mol}$--$M_\star$ relation derived from the molecular depletion times, $\tau_{\rm dep}=$ $M_{\rm mol}/{\rm SFR}$, for xGASS-CO MS galaxies by \cite{Saintonge&Catinella2022}. The shaded-green area is the ${\rm SFE}_{\rm mol}(r<R_{\rm e})$ range covered by EDGE-CALIFA spiral galaxies within $1\sigma$ scatter. {\it Right:} ${\rm SFE}_{\rm mol}(r<R_{\rm e})$ vs \hi-Groups. Conventions are as in left panel of Fig. \ref{Rmol_atom_star_vs_HIClass}. The results shown in Figs. \ref{SFEmol_vs_Rgal_2} and \ref{SFEmol_vs_Rgal} suggest that the systematic decrease of ${\rm SFE}_{\rm mol}(r<R_{\rm e})$ is a consequence of changes to the state of the gas or the star-formation process caused by the source of the morpho-kinematic perturbations that affect the \hi\ in the outer disks of VERTICO galaxies.}
\label{SFEmol_vs_Rgal}
\end{figure*}

\begin{figure}
\hspace{-0.3cm}
\includegraphics[width=9.5cm]{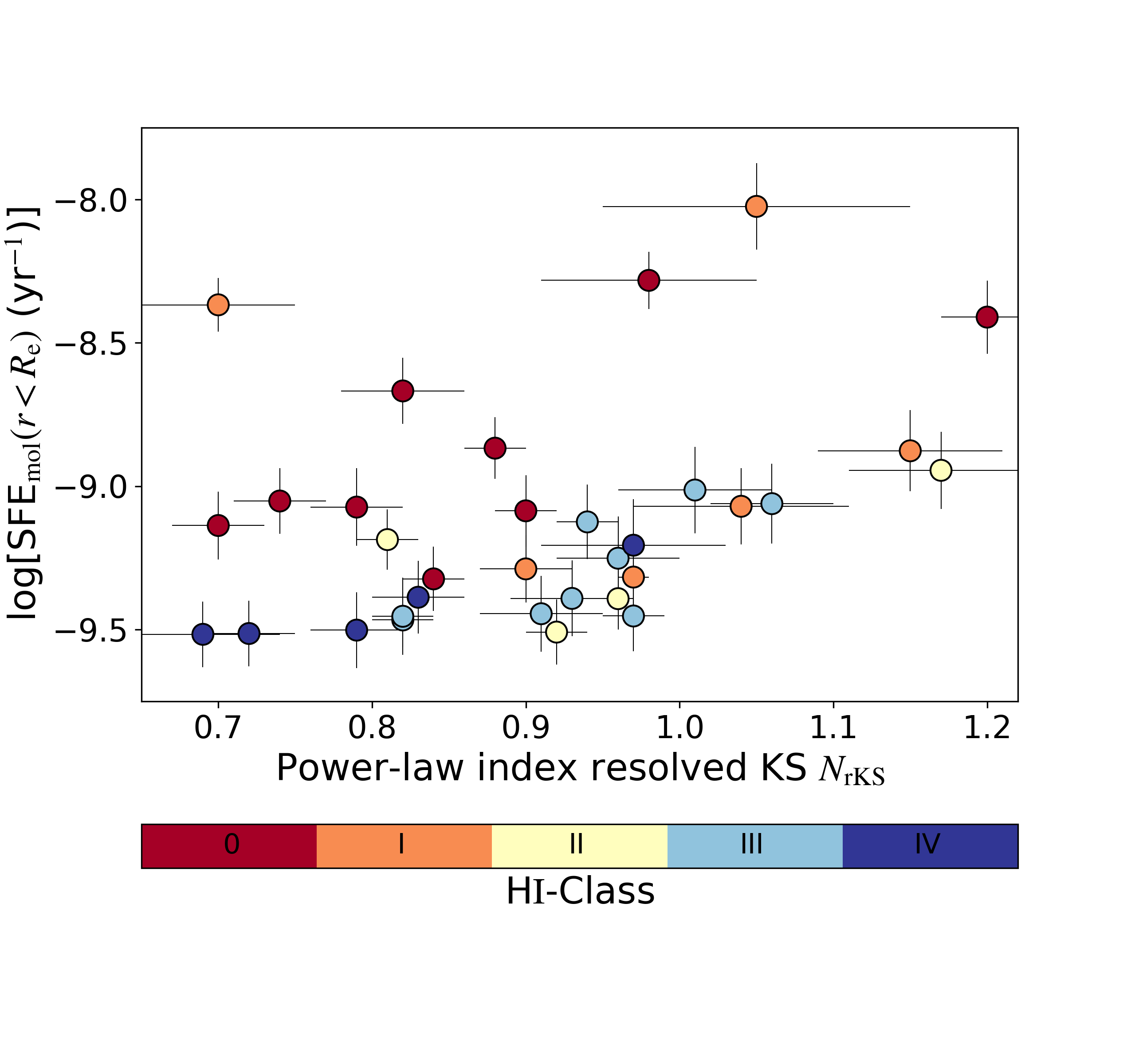}  
\caption{The star formation efficiency of the molecular mass within $R_{\rm e}$, SFE$_{\rm mol}(r<R_{\rm e})$, vs the best-fit power-law index of the resolved Kennicut-Schmidt, $N_{\rm rKS}$, from Jimenez-Donaire et al. (submitted). Although without a significant correlation between SFE$_{\rm mol}(r<R_{\rm e})$ and $N_{\rm rKS}$ (Pearson r$_{\rm p}=0.3$), \hi-Classes III and IV (blue circles) seem to group at both lower $N_{\rm rKS}$ and ${\rm SFE}_{\rm mol}(r<R_{\rm e})$ than \hi-Classes 0 and I (red circles), and vice versa.}
\label{SFE_Nks}
\end{figure}
       
\subsubsection{SFE vs Radius and Environment}  
\label{SFEmol_radius}

Figure \ref{SFEmol_vs_Rgal_2} shows the molecular gas mass within $R_{\rm e}$, $M_{\rm mol}(r<R_{\rm e})$ (top panel), and the star formation rate within $R_{\rm e}$, SFR$(r<R_{\rm e})$ (bottom panel), vs \hi-Class. Note that $M_{\rm mol}(r<R_{\rm e})$ remains almost constant with \hi-Class; on average, $M_{\rm mol}(r<R_{\rm e})$ values for VERTICO galaxies are similar to the mean for EDGE-CALIFA spirals. Although the mean SFR$(r<R_{\rm e})$ does not vary significantly for \hi-Classes 0 and I, we note a systematic decrease of the SFR with increasing \hi-Class for Classes III to IV. The results shown in Figure \ref{SFEmol_vs_Rgal_2} are consistent with the expected difference in the total molecular gas masses and star formation rates between galaxies from VERTICO and EDGE-CALIFA surveys. While the former includes galaxies with lower $M_{\rm mol}$s than xGASS-CO main sequence galaxies, as shown in Table \ref{Table_2}, the latter encompasses actively star-forming galaxies and rich in molecular gas (see \citealt{Bolatto2017} for more details of the sample selection), hence with higher molecular gas masses than main sequence galaxies. 

The left panel of Figure \ref{SFEmol_vs_Rgal} shows the spatially resolved SFE$_{\rm mol}=\Sigma_{\rm SFR}/\Sigma_{\rm mol}$ as a function of galactocentric radius and colored by \hi-Class. We do not observe a statistically significant correlation between the efficiencies and radius. We note that while higher \hi-Classes have SFE$_{\rm mol}$ significantly below to the almost constant values for EDGE-CALIFA spirals, lower \hi-Classes tend to be in fair agreement with the latter and scattered to much larger values (likely due to variations in $\Sigma_{\rm SFR}$ values within these \hi-Classes; see Fig. \ref{Radial_profiles}). On average, lower \hi-Classes tend to have higher SFE$_{\rm mol}$ than upper \hi-Classes for the $r_{\rm gal}$ range covered here. The middle and right panels of Figure \ref{SFEmol_vs_Rgal} show the star formation efficiency of the molecular gas within $R_{\rm e}$, ${\rm SFE}_{\rm mol}(r<R_{\rm e}) = {\rm SFR} (r< R_{\rm e}) / M_{\rm mol} (r< R_{\rm e})$, as a function of \hi-Class. We note a systematic decrease of SFE$_{\rm mol}(r<R_{\rm e})$ with \hi-Class. Although SFE$_{\rm mol}(r<R_{\rm e})$ values are mostly within the range covered by the control samples for \hi-Classes 0 and I, \hi-Classes II--IV have notably lower efficiencies ($\sim0.3-0.5$ orders of magnitude, or $2-3$ times lower) than EDGE-CALIFA spirals (green shaded area) and xGASS-CO MS galaxies (black horizontal bars). We test how our efficiencies depend on the adopted $\alpha_{\rm CO}$ prescription. Although SFE$_{\rm mol}$ values increase by $\sim1.1-1.3$ dex when using $\alpha_{\rm CO}(\Sigma_{\rm total})$ in the region within $\sim0.6 r_{25}$, the SFE$_{\rm mol}$ trend does not vary significantly compared to that for $\alpha_{\rm CO,MW}$. Likewise, SFE$_{\rm mol}(r<R_{\rm e})$ still shows a similar systematical decrease with \hi-Class when adopting the two $\alpha_{\rm CO}$ prescriptions. These results suggest that the systematic decrease of ${\rm SFE}_{\rm mol}(r<R_{\rm e})$ seen in Figure \ref{SFEmol_vs_Rgal} is a consequence of changes to the star formation process caused by the source of the perturbations that affect the \hi\ in the external regions of the disk.

It is interesting to compare our results with other galaxy-scale indicators of the star formation efficiency in VERTICO galaxies. For instance, Jim\'enez-Donaire et al. (submitted) compute the best-fit power-law index of the resolved Kennicutt--Schmidt law, $N_{\rm rKS}$, based on the resolved scaling relations between $\Sigma_{\rm SFR}$ and $\Sigma_{\rm mol}$. Out of the 36 VERTICO galaxies with inclinations $i<80^\circ$ included in Jim\'enez-Donaire et al. (submitted), Figure \ref{SFE_Nks} shows the SFE$_{\rm mol}(r<R_{\rm e})$ vs $N_{\rm rKS}$ for 34 VERTICO galaxies for which they find $N_{\rm rKS}$ values statistically significant. We do not find a significant correlation between $N_{\rm rKS}$ and SFE$_{\rm mol}(r<R_{\rm e})$. However, in observing the distributions of $N_{\rm rKS}$ colored by \hi-Class, we note that \hi-Classes III and IV (blue circles) tend to group at both lower $N_{\rm rKS}$ and ${\rm SFE}_{\rm mol}(r<R_{\rm e})$ than \hi-Classes 0 and I (red circles), and vice versa. Since the resolved Kennicutt--Schmidt law index quantifies changes in the molecular gas efficiencies through $\Sigma_{\rm mol}$ and $\Sigma_{\rm SFR}$, lower efficiencies at the centers of \hi-Classes II-IV reflect locally low $\Sigma_{\rm SFR}/\Sigma_{\rm mol}$ ratios, which drive $N_{\rm rKS}$ values below unity. Jim\'enez-Donaire et al. (submitted) also show that, on average, the distribution of $N_{\rm rKS}$ in VERTICO galaxies peaks at lower values when compared to those for HERACLES \citep[][]{Leroy2013} and PHANGS \cite[][]{Pessa2021}. Our results are consistent with Jim\'enez-Donaire et al. (submitted), who suggest that their sub-linear $N_{\rm rKS}$ values found in most of the VERTICO galaxies indicate a decrease in the efficiency of the molecular gas for regions with high $\Sigma_{\rm mol}$ and reflects the broad variety of physical conditions present in Virgo galaxies.

Our results show that VERTICO galaxies tend to be less efficient at converting molecular gas into stars when their atomic gas is strongly affected by environmental mechanisms, at least in the region within $R_{\rm e}$. Analyzing 98 galaxies selected from the JCMT NGL survey, \cite{Mok2016} show that Virgo galaxies have longer molecular gas depletion times, $\tau_{\rm dep}=M_{\rm H_2}/$SFR$ = $SFE$^{-1}_{\rm H_2}$, when compared to group galaxies selected from a sample of 485 local galaxies included in \cite{Garcia1993}. They attribute this difference to a combination of environmental factors that increase the H$_2$ production and a decrease in the SFR in the presence of large amounts of molecular gas, which may reflect heating processes in the cluster environment or differences in the turbulent pressure. \cite{Lee2017} find an increase in the CO surface brightness (an increase in $\Sigma_{\rm H_2}$) in a region close to the galactic center subjected to intense ram pressure in the Virgo galaxy NGC 4402 (also confirmed by \citealt{Cramer2020}), which seems to be tied to bright FUV and H$\alpha$ emission associated with intense star formation. \cite{Zabel2020} also observe an enhancement in the H$_2$ star formation efficiencies, SFE$=\Sigma_{\rm SFR}/\Sigma_{\rm H_2}$, of Fornax cluster galaxies (particularly at low masses) for galaxies on first passage through the cluster. They suggest that these changes might be driven by environmental mechanisms (e.g., RPS or tidal interactions). \cite{Morokuma-Matsui2021} analyze the atomic and molecular gas properties of massive Virgo galaxies ($M_\star > 10^{9} M_\odot$), which are selected from the Extended Virgo Cluster Catalog \citep[EVCC;][]{Kim_2014}, within $r<3R_{200}$ ($R_{200}$ is the radius where the mean interior density is 200 times the critical density of the Universe). They find that Virgo galaxies have lower SFRs and higher SFE$_{\rm H_2} = $SFR$/M_{\rm H_2}$ compared to field galaxies with offsets from the main sequence of the star-forming galaxies $\Delta$(MS)$<0.0$. In addition, they note that Virgo galaxies have both lower gas fractions ($M_{\rm H_2}/M_{\star}$ and $M_{\rm HI}/M_{\star}$) and higher SFEs compared to field galaxies.  \cite{Roberts2022} also find evidence of enhanced star formation on the leading side of four identified jellyfish galaxies selected from Perseus cluster using radio LOw Frequency ARray (LOFAR) continuum at 144 MHZ and H$\alpha$ data. They find a positive correlation between  H$\alpha$+[NII] surface brightness and the orientation angle of sources with respect to the stripped tail, which is consistent with gas compression (i.e., the increasing of the star production) induced by ram pressure along the interface between the ICM and the galaxy. \cite{Lee2022} analyze ALMA ACA $^{12}$CO(J=1-0) and \hi\ data for 31 galaxies selected from the Group Evolution Multiwavelength Study survey \citep[GEMS; ][]{Osmond&Ponman2004,Forbes2006}, finding that some members with highly asymmetric morphologies in CO and \hi\ images (e.g., driven by tidal interactions and RPS) have a significant decrease in both SFR (e.g., due to gas becoming stable against gravitational collapse), and gas fractions, suggesting a decrease of $\Sigma_{\rm mol}$ due to the suppression of the \hi-to-H$_2$ conversion. Numerical simulations have found that the star formation activity is generally amplified in galaxy centers by ram pressure \citep[e.g.,][]{Tonnesen&Bryan2012,Bekki2014}; particularly, some of them have shown that the star production can be boosted due to gas compression at the edges of the disks \citep[e.g., ][]{Roediger2014,Boselli2021}.

Since our integrated efficiencies are computed within $R_{\rm e}$, it is possible that the systematic decrease of SFE$_{\rm mol}(r<R_{\rm e})$ with \hi-Class could be in part caused by morphological quenching, MQ, if most gas is driven into bulge-dominated galaxy centers. MQ (\citealt{Martig2009}) is able to produce a gravitational stabilization of the gas within the bulge region, preventing the fragmentation into bound star-forming clumps. Numerical simulations performed by \cite{Gensior2020} show that spheroids drive turbulence and increase the gas velocity dispersion, virial parameter, and turbulent gas pressure towards the galaxy centers, which are mostly dependent on the bulge mass ($M_{\rm b}$). They note that the more massive the bulges are, the higher the level of turbulence. Therefore, the stellar spheroid stabilizes the ISM of the host galaxy by increasing the shear velocity and the gas velocity dispersion, thus preventing the gravitational instability of the gas reservoirs and suppressing the fragmentation responsible for the disruption of the ISM by stellar feedback. MQ has been shown to potentially operate not only on early-type massive galaxies with a strong bulge component, but it can also work at any $M_\star$ range \citep[e.g.,][]{Catalan-Torrecilla2015,Catalan-Torrecilla2017}. However, some studies \citep[e.g.,][]{Cook2019,Cook2020} have noted that the regulation of the SFR in main sequence galaxies is more related to physical processes acting on the disk rather than the contribution from bulges.

\begin{figure}
\hspace{-0.3cm}
\includegraphics[width=9.5cm]{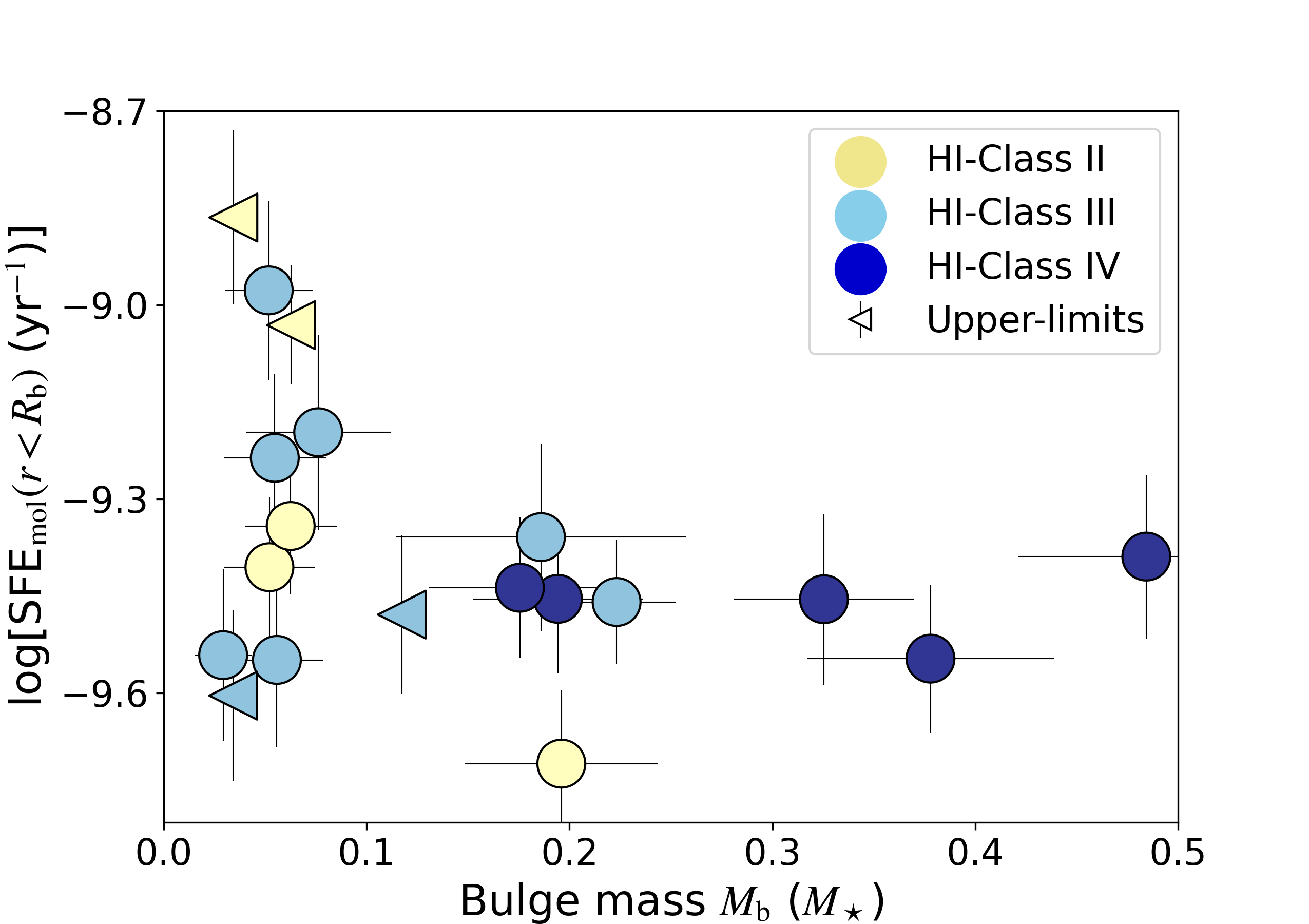}  
\caption{The star formation efficiency of the molecular mass within the radius of the bulge $R_{\rm b}$, SFE$_{\rm mol}(r<R_{\rm b})$, vs the bulge mass $M_{\rm b}$ (in units of the total stellar mass) for \hi-Classes II, III, and IV. Symbols are colorcoded by \hi-Class as in Figure \ref{SFE_Nks}. The horizontal arrows are upper-limits for $M_{\rm b}$ since no clear breaks in the stellar radial profiles due to bulges are identified; therefore, $M_{\rm b}$ in these cases is the mass enclosed within $0.1 r_{\rm 25}$, which corresponds to $r_{\rm gal}\approx 1.0$ kpc at the Virgo cluster distance ($16.5$ Mpc; \citealt{Mei2007}).}
\label{SFE_Mb}
\end{figure}

Figure \ref{SFE_Mb} shows the star formation efficiency of the molecular mass within the radius of the bulge $R_{\rm b}$, SFE$_{\rm mol}(r<R_{\rm b})$, vs bulge mass, $M_{\rm b}$, for \hi-Classes II, III, and IV, since these are the \hi-Classes with more central concentration of the molecular gas (see Fig. \ref{3_Exp_lenght}). We compute $M_{\rm b}$ by visually inspecting the $\Sigma_\star$ radial profiles (orange-solid line in Fig. \ref{Radial_profiles}) in the region within $R_{\rm e}$. Then we identify the $r_{\rm gal}$ where there is a break with respect to the stellar exponential profiles (red-dashed line in Fig. \ref{Radial_profiles}). We use this break to define $R_{\rm b}$. In cases where this break is not evident, we take $R_{\rm b}=0.1 r_{25}$ (similar to the fiducial radius of $1.0$ kpc of the bulge adopted by \citealt{Regan2001}), which matches the physical resolution in VERTICO CO data at the Virgo cluster distance of $16.5$ Mpc (\citealt{Mei2007}), to compute a conservative upper-limit on $M_{\rm b}$ (horizontal arrows). Finally, we integrate the $\Sigma_\star$ radial profiles between $0$ and $R_{\rm b}$ to obtain $M_{\rm b}$. As seen in Figure \ref{SFE_Mb}, we do not find a statistically significant anticorrelation between SFE$_{\rm mol}(r<R_{\rm b})$ and $M_{\rm b}$ (Pearson $\rm r_p =-0.3$; $p$-value$<0.7$). Since we are using a simple approach to compute $R_{\rm b}$, smaller $M_{\rm b}$ are succeptible to be not reliable estimations (e.g., the ``breaks'' in $\Sigma_{\star}$ could be due other circumnuclear structures). However, the figure suggests that VERTICO galaxies with more centrally concentrated molecular gas (specifically in the higher \hi-Classes) tend to be less efficient at converting the H$_2$ into stars when they host a more massive stellar bulge. 

If the cluster environmental mechanisms act by pushing the molecular gas to central regions within the influence of the bulge, MQ is then a good candidate to explain the SFE$_{\rm mol}(r<R_{\rm e})$ decrease observed in higher \hi-Classes via gas stabilization (at least in some VERTICO galaxies). Detailed studies of the molecular gas dynamics within the bulge region in Virgo galaxies are therefore important to establish the actual connection between the SFE$_{\rm mol}$, the \hi-Class, and the gravitational stability of the gas.

\section{Summary and conclusions}
\label{S6_Conclusion}

We present a study of the molecular-to-atomic gas ratio, the star formation efficiency, and their dependence on other physical parameters in 38 galaxies selected from the VERTICO survey. We analyze $^{12}$CO($J$=2--1) datacubes with ${9}\arcsec$ angular resolution (except for the ${13}\arcsec$ NGC 4321 datacube) at 10 km s$^{-1}$ channel width, along with \hi\, velocities extracted from VIVA survey. We implement spectral stacking of CO spectra to co-add them coherently by using \hi\ velocities from the VIVA survey, and  measure $\Sigma_{\rm mol}$ out to typical galactocentric radii of $r\approx1.2\,r_{25}$ by coherently averaging (stacking) spectra in elliptical annuli using the HI velocity as a reference in each pixel. We use a constant Milky Way CO-to-H$_2$ conversion factor prescription $\alpha_{\rm CO, MW} = 4.3$ M$_\odot$ $[\rm K \, km \, s^{-1} \, pc^{2}]^{-1}$ (\citealt{Walter2008}), and a line Rayleigh-Jeans brightness temperature ratio of $R_{21}=I_{\rm CO(2-1)}/I_{\rm CO(1-0)}\sim0.77$. Although the adoption of a variable $\alpha_{\rm CO}$ has some impact on the molecular surface densities, it does not change the trends that we have found in this work. We warn that our exploration of the effects of $\alpha_{\rm CO}$ dependence is limited and it deserves a more careful analysis in future VERTICO projects. We perform a systematic analysis to explore molecular disk sizes, molecular-to-atomic gas ratios, and the star formation efficiency of the molecular gas in VERTICO in comparison to field samples. Our main conclusions are as follows:

\begin{enumerate}
    \item We determine the molecular and stellar exponential disk scale lengths, $l_{\rm mol}$ and $l_{\star}$, respectively,  by fitting an exponential function to the $\Sigma_{\rm mol}$ and $\Sigma_{\star}$ radial profiles for 33 VERTICO galaxies (see Fig. \ref{3_Exp_lenght}). We find that $l_{\rm mol} = (0.62\pm0.05) \, l_{\star}$ ($\sim$3:5 relation). When compared with the equivalent relation observed for galaxies selected from the field (e.g., EDGE-CALIFA survey; \citealt{Villanueva2021}), the CO emission in VERTICO galaxies is more centrally concentrated than the stellar surface density. Moreover, galaxies with a stronger degree of H\,{\sc i} perturbation \citep[as classified by][]{Yoon2017} tentatively show more compact CO distributions (this is particularly true for \hi-Class III galaxies).
	
	\item To test how the Virgo environment may be affecting the atomic-to-molecular gas transition, we compute molecular-to-stellar and atomic-to-stellar ratios as a function of galactocentric radius (left and middle panels of Fig. \ref{Rmol_atom_star_vs_Rgal}). To control for the effects of stellar mass when comparing to a field sample, we use the results obtained from xGASS-CO for stellar masses matched to each subsample in VERTICO that has been broken into \citet{Yoon2017} \hi-Classes. While the molecular-to-stellar mass ratio integrated out to $R_{\rm e}$ in VERTICO galaxies is completely consistent with the xGASS-CO sample, the atomic-to-stellar mass ratio integrated out to $R_{\rm e}$ shows a systematic decrease with increasing \hi-Class, reflecting the known significant \hi\ deficiencies in the high \hi-Classes \citep{Yoon2017}. 
	
	\item The resolved molecular-to-atomic gas ratio, $R_{\rm mol}$, decreases with increasing galactocentric radius (left panel of Fig. \ref{Rmol_vs_Rgal}) for all \hi-Classes, as expected in galaxies. However, we find a systematic increase in the molecular-to-atomic gas ratio integrated out to $R_{\rm e}$ with increasing \hi-Class (right panel of Fig. \ref{Rmol_vs_Rgal}). Together with the previous point, these results suggest that although environmental processes act on the atomic and the molecular gas simultaneously, the atomic gas content is considerably more affected than the molecular gas content. Consequently, the morpho-kinematic \hi\, features of VERTICO galaxies are not a good predictor for their molecular gas content.
	
	\item There is a dependency of the star formation efficiency of the molecular gas within $R_{\rm e}$ on the morpho-kinematic \hi\, features in VERTICO galaxies, but no strong systematic trends with galactocentric radius (Fig. \ref{SFEmol_vs_Rgal}). On average, VERTICO galaxies tend to be decreasingly efficient at converting their molecular gas into stars when their atomic gas is strongly perturbed by environmental effects. Although we do not find a statistically significant correlation between star formation efficiency within the bulge radius and the mass of the bulge we observe that galaxies with more centrally concentrated molecular gas tend to be less efficient at converting their H$_2$ into stars when they host a more massive stellar bulge. 
    
\end{enumerate}

The general picture is that both the molecular and the atomic gas are affected by the Virgo environment, but in different manners (see also \citealt{Cortese2021} and references therein). First, the mechanisms that remove \hi\ in the cluster do not seem to significantly remove molecular gas. Instead, they appear to drive the molecular component toward the central regions, resulting in molecular disks with shorter scale lengths than their companion stellar disks. The removal of atomic gas results in galaxies that have high molecular-to-atomic ratios. However, these more centrally concentrated molecular regions with higher molecular-to-atomic ratios exhibit lower molecular star formation efficiencies than observed in field galaxies.  A different (but also complementary) explanation is the removal of the molecular gas (e.g. by RPS acting preferentially in the more diffuse H$_2$ that is not strongly tied to GMCs) in combination with a simultaneous phase transition from \hi-to-H$_2$ in the inner part of the galaxy triggered by the mechanisms that remove \hi. The molecular gas removal also agrees with H$_2$ observations in the tails of jellyfish galaxies \citep[e.g., ][]{Moretti2020}, which could otherwise be explained as a phase transition from \hi-to-H$_2$ in the tail itself. 

Future projects may complement the $^{12}$CO(J=2-1) observations presented here with more accurate $\alpha_{\rm CO}$ prescriptions (e.g., including metallicity indicators in Equation \ref{eq:alpha_co}), the impact of the environment on the diffuse gas, or sub-kpc-scale CO observations within the bulge regions of high \hi-Class galaxies. Also, detailed kinematic analyses are required to test the potential impact of environmental (or intrinsic) effects on the molecular gas stability, which may be decreasing the SFE in VERTICO galaxies.

\section{acknowledgments}

V. V. acknowledges support from the scholarship ANID-FULBRIGHT BIO 2016 - 56160020 and funding from NRAO Student Observing Support (SOS) - SOSPA7-014. A. D. B., S. V., and V. V., acknowledge partial support from NSF-AST2108140. T.B. acknowledges support from the National Research Council of Canada via the Plaskett Fellowship of the Dominion Astrophysical Observatory. A.R.H.S. gratefully acknowledges funding through the Jim Buckee Fellowship at ICRAR-UWA. T.A.D. acknowledges support from the UK Science and Technology Facilities Council through grants ST/S00033X/1 and ST/W000830/1. B. L. acknowledges the support from the Korea Astronomy and Space Science Institute grant funded by the Korea government (MSIT) (Project No. 2022-1- 840-05). L.C.P. acknowledges support from the Natural Sciences and Engineering Research Council of Canada. A.C. acknowledges support from the NRF (grant No. 2022R1A2C100298211 and 2022R1A6A1A03053472) by the Korean government. L.C. acknowledges support from the Australian Research Council Discovery Project and Future Fellowship funding schemes (DP210100337, FT180100066).

This paper makes use of the following ALMA data:

\begin{itemize}
\setlength{\itemsep}{0pt}
\setlength{\parskip}{0pt}
\setlength{\parsep}{0pt}
    \item ADS/JAO.ALMA \#2019.1.00763.L,
    \item ADS/JAO.ALMA \#2017.1.00886.L,
    \item ADS/JAO.ALMA \#2016.1.00912.S,
    \item ADS/JAO.ALMA \#2015.1.00956.S.
 \end{itemize}

ALMA is a partnership of ESO (representing its member states), NSF (USA) and NINS (Japan), together with NRC (Canada), MOST and ASIAA (Taiwan), and KASI (Republic of Korea), in cooperation with the Republic of Chile. The Joint ALMA Observatory is operated by ESO, AUI/NRAO and NAOJ. In addition, publications from NA authors must include the standard NRAO acknowledgement: The National Radio Astronomy Observatory is a facility of the National Science Foundation operated under cooperative agreement by Associated Universities, Inc. 
Part of this work was conducted on the unceded territory of the Lekwungen and Coast Salish peoples. We acknowledge and respect the Songhees, Esquimalt, WS\'{A}NE\'{C} and T'Sou-ke Nations whose historical relationships with the land continue to this day.
Support for CARMA construction was derived from the Gordon and Betty Moore Foundation, the Kenneth T. and Eileen L. Norris Foundation, the James S. McDonnell Foundation, the Associates of the California Institute of Technology, the University of Chicago, the states of California, Illinois, and Maryland, and the NSF. CARMA development and operations were supported by the NSF under a cooperative agreement and by the CARMA partner universities. This research is based on observations collected at the Centro Astron\'{o}mico Hispano-Alem\'{a}n (CAHA) at Calar Alto, operated jointly by the Max-Planck Institut f\"{u}r Astronomie (MPA) and the Instituto de Astrofisica de Andalucia (CSIC).

This research has made use of the NASA/IPAC Extragalactic Database (NED), which is operated by the Jet Propulsion Laboratory, California Institute of Technology, under contract with the National Aeronautics and Space Administration.

Parts of this research were conducted by the Australian Research Council Centre of Excellence for All Sky Astrophysics in 3 Dimensions (ASTRO 3D), through project number CE170100013.

\software{ Astropy \citep{AstropyCollaboration2018}, MatPlotLib \citep{Hunter2007},
NumPy \citep{Harris2020}, SciPy \citep{2020SciPy-NMeth}, seaborn \citep{Waskom2021}, Scikit-learn \citep{scikit-learn}.}

\bibliography{main}{}
\bibliographystyle{aasjournal}



\end{document}